\documentclass[a4paper,usenatbib]{mnras}

\makeatletter

\newcommand{\Rmnum}[1]{\expandafter\@slowromancap\romannumeral #1@}
\newcommand{\lsim}{\lesssim}
\newcommand{\gsim}{\lower0.6ex\vbox{\hbox{$\buildrel{\textstyle >}\over{\sim}\ $}}}

\makeatother

\usepackage{newtxtext,newtxmath}
\usepackage{graphicx}	
\usepackage{amsmath}	
\usepackage{subcaption}
\usepackage{booktabs}
\usepackage[dvipsnames,svgnames]{xcolor}
\usepackage{color}
\newcommand{\yueying}[1]{\textcolor{black}{#1}}

\def\hmpc{h^{-1}{\rm Mpc}}
\def\hgpc{\;h^{-1}{\rm Gpc}}
\def\invhmpc{\;h\;{\rm Mpc}^{-1}}

\def\hkpc{h^{-1}\, {\rm kpc}}

\def\msun{\, M_{\odot}}

\title[Super-resolution simulations]
{AI-assisted super-resolution cosmological simulations II: Halo substructures, velocities and higher order statistics}

\author[Y. Ni et al.]
{Yueying Ni$^{1,2}$\thanks{Email:yueyingn@andrew.cmu.edu}, Yin Li$^{3}$, Patrick Lachance$^{1,2}$,
Rupert A.~C. Croft$^{1,2}$, Tiziana Di Matteo$^{1,2}$, Simeon Bird$^{4}$,  
\newauthor Yu Feng$^{5}$ \\
$^1$ McWilliams Center for Cosmology, Department of Physics, Carnegie Mellon University, Pittsburgh, PA 15213 \\
$^2$ NSF AI Planning Institute for Physics of the Future, 
Carnegie   Mellon  University, Pittsburgh, PA 15213, USA \\
$^3$ Center for Computational Astrophysics \& Center for Computational Mathematics, Flatiron Institute, 162 5th Avenue, New York, NY 10010 \\
$^4$ Department of Physics and Astronomy, University of California Riverside,
900 University Ave, Riverside, CA 92521 \\
$^5$ Berkeley Center for Cosmological Physics and Department of Physics, University of California, Berkeley, CA 94720, USA \\
}

\date{Accepted XXX. Received YYY; in original form ZZZ}

\pubyear{2021}
\begin{document}
\maketitle

\begin{abstract}
In this work, we expand and test the capabilities of our recently developed super-resolution (SR) model to generate high-resolution (HR) realizations of the full phase-space
matter distribution, including both displacement and velocity,
from computationally cheap low-resolution (LR) cosmological N-body simulations.
The SR model enhances the simulation resolution by generating 512 times more tracer particles, extending into the deeply non-linear regime where complex structure formation processes take place.
We validate the SR model by deploying the model in 10 test simulations of box size $100 \, \hmpc$, and examine the matter power spectra, bispectra and 2D power spectra in redshift space.
We find the generated SR field matches the true HR result at percent level down to scales of $k \sim 10 \invhmpc$.
We also identify and inspect dark matter halos and their substructures.
Our SR model generates visually authentic small-scale structures, that cannot be resolved by the LR input, and are in good statistical agreement with the real HR results. 
The SR model performs satisfactorily on the halo occupation distribution, halo correlations in both real and redshift space, and the pairwise velocity distribution, matching the HR results with comparable scatter, thus demonstrating its potential in making mock halo catalogs.
The SR technique can be a powerful and promising tool for modelling  small-scale galaxy formation physics in large cosmological volumes.
\end{abstract}


\section{Introduction}
\label{section1:introduction}

Cosmological simulations have steadily increased in size and
complexity over the last 40 years (see e.g., \citealt{vogelsberger20} and references
therein). In a cosmological volume
(a fair sample of the universe), galaxies with masses as small as $10^{8} \msun$
can now
be simulated with (baryonic) mass resolutions of $10^{4}-10^{5} \msun$, resolving
length scales approaching and below 300 pc (e.g., \citealt{nelson20}, \citealt{khim20} ).
Progress in this area has been closely linked to developments in
high performance computing (such as the advent of parallel programming, e.g., \citealt{salmon91}),
and algorithms (including trees for gravity, \citealt{barnes86}, and adaptive time-stepping,
\citealt{porter85}). Nevertheless, with foreseeable improvements in both areas, cosmological
simulations which include  star and planet formation seem likely to be out of
reach for many years \citep{nagamine18}.
However, the Artificial Intelligence (AI) revolution now touching many
aspects of society offers us a way to make advances, and potentially
reach many of the scientific goals that raw increases of computing power by
many orders of magnitude would otherwise necessitate.
Simulations of astrophysical processes is a multi-scale problem, and different techniques are used in scales below a cosmological context.
For example the recent
STARFORGE models of magnetohydrodynamic star formation \citep{grudi20}
form clusters of stars with typical mass resolution
 of $10^{-3}\msun$. Formation of Earth-like
planets in a dissipating gas disk was modelled by
\cite{walsh19}. 
Traditionally, such small scale physical processes are included in cosmological simulations as sub-grid effective or mean-field models, in order to control the computation cost and the complexity of the numerical model.
However, to fully understand the roles of these astrophysical processes in cosmology, 
one can ask whether it is possible to
include them in a consistent fashion.

The technique of AI-assisted super-resolution offers us a route towards
doing this. Super-resolution (SR) is the addition or recovery of
information below the resolution scale, and it is most employed in the context of two dimensional images. 
Among the variety of methods for carrying out SR,
AI assistance has shown great promise \citep{Yue2016}, and is the basis of our work on this topic.

SR enhancement is a very challenging problem even in the context of two dimensional images, because 
there are always multiple high resolution (HR) realizations corresponding to a single low resolution (LR) image. 
In spite of this, however, new techniques based on Deep Learning (DL) have
proven amazingly effective \citep[see e.g., the review of DL based image super-resolution by][]{wang2020}.  
One of the most promising DL approaches so far to SR use Generative Adversarial Nets (GAN)
\citep[e.g. SRGAN;][]{Ledig2016}.
GAN \citep{goodfellow2014generative} is a class of DL system in which two Neural Networks (NN) compete in a game: the generator network generates candidates while the discriminator network evaluates the quality of the candidates by comparing learned `features` or statistics against the authentic high resolution images.
As the networks are trained against examples of training data, the fidelity of the output increases as both the generator and the discriminator become better at doing their jobs. 

Deep Learning is finding many other applications in training surrogate models for cosmological simulations.  
For example, NNs have been used to predict the nonlinear structure formation from the linear cosmological initial condition~\citep{he2019learning, berger2019, Bernardini2020, Renan20}
and from the dark matter density field~\citep{ramanah2019painting}.
\yueying{Generative networks have been applied to directly produce multiple cosmological simulation outputs such as 2D images \citep{Rodrguez2018}, density fields \citep{Perraudin2019} and cosmological mass maps \citep{Perraudin2020}.}
Models have also been trained to predict various baryonic properties from dark matter only simulation, such as galaxy distribution \citep{Modi2018, zhang2019}, thermal Sunyaev-Zeldovich (tSZ) effect \citep{Troster2019}, 21 cm emission from neutral hydrogen \citep{wadekar2020hinet}, stellar maps and various gas properties \citep{Dai2021}, etc.
Recently, \citet{Villanavarro2020b, Villanavarro2020a} have started the Cosmology and Astrophysics with MachinE Learning Simulations (CAMELS), a set of over 4000 hydrodynamical simulations run with different hydrodynamic solvers and subgrid models for galaxy formation, providing a large training set to study  baryonic effects with machine learning applications.
Works has also been carried out to apply SR technique to directly enhance the spatial or mass resolution of cosmological simulations.
\cite{KodiRamanah2020} developed a SR network to map density fields of LR cosmological simulations to that of the HR ones.

In our previous work \cite{Li2021} (hereafter Paper I), we presented a powerful SR model for cosmological simulations. 
The model enhances the resolution of cosmological $N$-body simulation (so far, of dark matter only) by a factor of 8 in spatial and 512 in mass, producing the full particle distribution in Lagrangian prescription.
Our SR model extends to scales orders of magnitude below the LR input and give statistically good match compared with the HR counterparts, such as matter power spectrum and halo mass function.

However, in this first work, we did not explore higher order statistics of the large-scale structure and the internal sub-structure of gravitationally bound objects. These statistics are more difficult to predict due to increased accuracy requirement on the modelling of nonlinearity.
We also did not include peculiar velocity field of the tracer particles when training our SR model.

In the present paper, we improve our SR method and compare these statistics of our SR model against the HR results.
Our motivation for the work in this paper is the following:
\begin{itemize}
    \item In addition to the 3-dimensional displacement field, we now also generate the SR enhanced 3-dimensional velocity field, expanding the model to cover the full 6-dimensional phase space. The 6-dimensional field can be analyzed in the same manner as a full $N$-body simulation;
    \item We make use of the velocities to test redshift-space clustering statistics;
    \item We test higher order statistics, specifically the bispectrum as a measurement of the non-Gaussianity in the evolved density field is captured in SR;
    \item We make the first attempts at predicting the spatial clustering as well as the internal substructure of dark matter halos in SR simulations.
\end{itemize}
We note that the last point (modelling substructure) is highly non-trivial with our Lagrangian coordinates based SR enhancement approach because it requires that the results must remain accurate well beyond orbit-crossing.

The paper is organized as follows. 
In Section~\ref{section2:Method}, we review our methodology, training dataset and the training process, as well as the GAN architecture used to train our generative model.
We present our major results in Section~\ref{section3:Result}, where Section~\ref{subsection:full-field} examines the statistics of the full density and velocity field generated by the SR model, while Section~\ref{subsection: halo-catalog} focuses on the statistics of mock halo catalogs.
We discuss our results in Section~\ref{section4:Discussion} and conclude in Section~\ref{section5:Conclusion}.
\section{Method}
\label{section2:Method}

\begin{figure*}
\centering
  \includegraphics[width=1.0\textwidth]{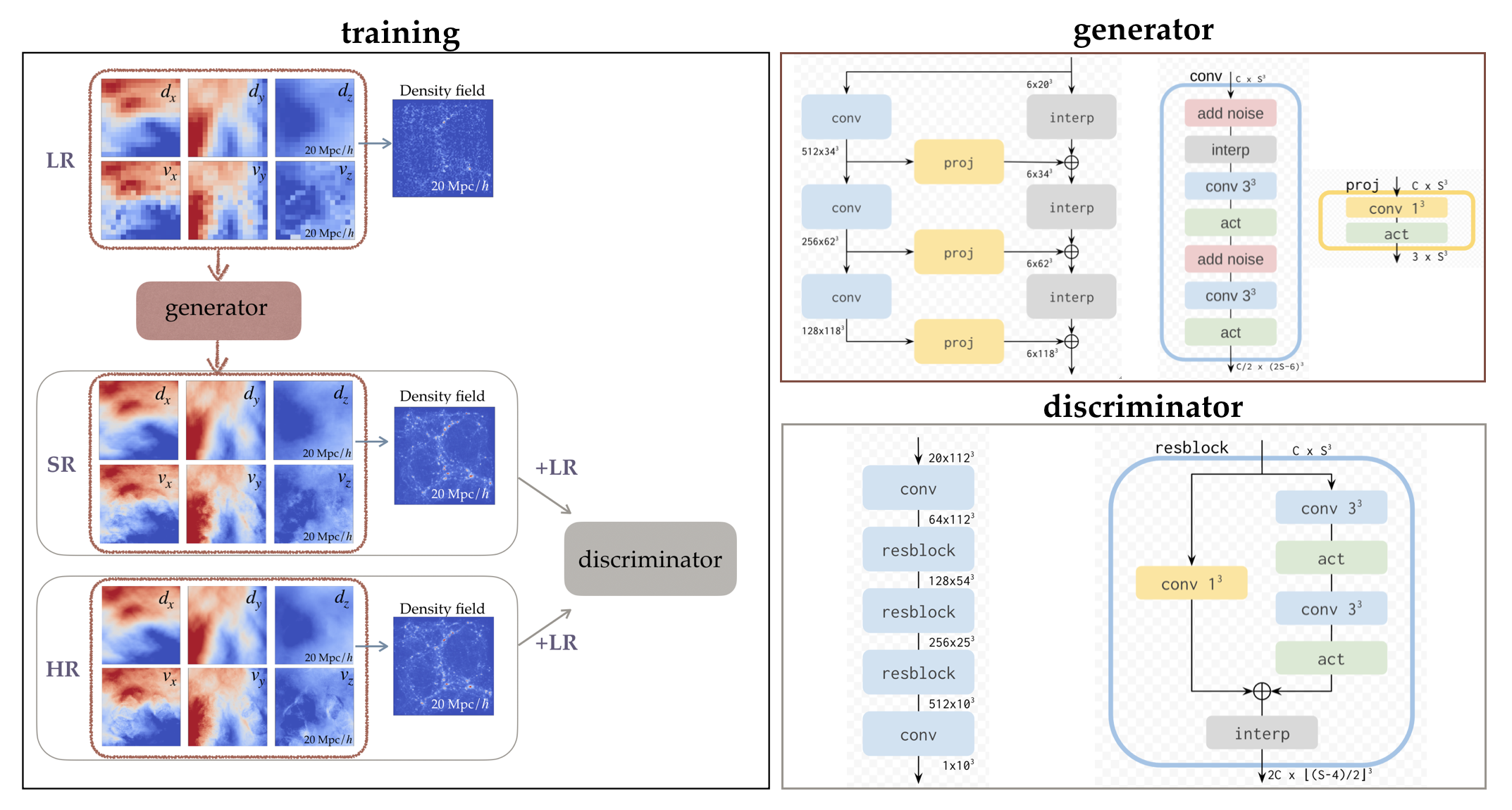}
  \caption{\textit{\textbf{Left panel:}}
  Schematic of our GAN training process.
  The generator takes as input 6 channels of displacement and velocity fields of the simulation particles (the Lagrangian description),
  and generates the super resolution fields.
  The discriminator scores the generated SR samples and their HR counterparts to assess their authenticity.
  In addition, we concatenate the low resolution input and Eulerian space density field (computed from displacement field using the CIC scheme) to the discriminator.
  \textit{\textbf{Right panel:}} Architecture of the generator and discriminator networks. \textit{\textbf{Generator:}} The left part shows the entire network, with details of the components given on the right. The ladder-shaped generator upsamples by a factor of two at each rung.
  The 3 convolution blocks on the left (“conv” in blue plates) operates in the high-dimensional latent space, and is projected (“proj” in yellow plates) at every step to the low-dimensional output space on the right rail.
  The projected results are then upsampled by linear interpolation (“interp” in gray plates), before being summed into the output.
  Noises (red plates) are added in the "conv" block to bring stochasticity to each resolution level. 
  \textit{\textbf{Discriminator:}} Composed from residual blocks with detailed structures given on the right.
  In addition to the residual blocks, the first ``conv'' block is a $1^3$ convolution followed by an activation, and the last one has a $1^3$ convolution to double the channel, an activation, and a $1^3$ convolution to reduce single channel.
  All activation functions (``act'' in green plates) in both generator and discriminator are Leaky ReLU with slope $0.2$ for negative values.
  }\label{fig:training}
\end{figure*}

\subsection{Super-resolving an $N$-body simulation}

$N$-body simulation is a powerful numerical method for solving the non-linear evolution of  cosmological structure formation.
By discretizing the matter distribution into mass particles and evolving them under gravity, it predicts the 6D phase-space distribution of the evolved dynamic field through the positions and velocities of a large number of massive tracer particles. 

The number of tracer particles $N$ determines the scale upon which the underlying physical field is properly resolved.
However, with most of the popular gravity solvers (e.g., the Tree-PM method, \citealt{bagla02}) for cosmological simulation, the computational complexity of $N$-body simulations typically scales with number of
time steps to integrate the particle equations of motion.
High resolution $N$-body simulations can be rather costly due to
accurate time integration of the nonlinear orbits, even with the help of adaptive time stepping schemes.

In this work, we train an SR model to generate HR $N$-body simulation outputs given LR $N$-body inputs. 
The generated SR field preserves the large scale features of the LR input, and add small scale structures in the density and velocity fields that statistically match the predictions of HR $N$-body simulations but are absent in the LR inputs.

We perform the SR simulation task following the Lagrangian description, where the particles are tracers of the displacement field $\mathbf{d}_i$ and the velocity field $\mathbf{v}_i$. 
The displacement field $\mathbf{d}_i = \mathbf{x}_i - \mathbf{q}_i$, where $\mathbf{x}_i$ is the current position of the tracer particle, and $\mathbf{q}_i$ is the original, unperturbed position of the particle (typically on a uniform grid as in our case). 
The $\{ \mathbf{d}_i, \mathbf{v}_i \}$ ($i = 1 ... N$) pair of Lagrangian fields can be concatenated and structured as a 3D image with 6 channels.
Each channel corresponds to one component of the displacement or velocity vector of the tracer particle originating from the $i$-th voxel.
The generative model takes the LR 6-channel field as input, and outputs a realization of the $\{ \mathbf{d}_i, \mathbf{v}_i \}$ field, representing more tracer particles of higher mass resolution. 
There are several advantages of using the Lagrangian description.
(i) By learning the displacement of tracer particles (instead of the density field in the Eulerian description), the mass of the output field is naturally conserved. 
(ii) The generated SR field can be formatted identically to the output from a real $N$-body simulation, with distinguishable tracer particles each evolved through time. We obtain the particle positions by moving them from their original positions on the lattice using the displacement vectors.
\yueying{(iii) The Lagrangian prescription preserves an advantage of  particle-based $N$-body simulations:  they adaptively resolve small scales in high-density regions.
As a result it may  better describe fields with a large dynamic range compared to the Eulerian prescription with the same grid size. 
In the Eulerian prescription, one needs to map a simulation field onto a uniform grid, where the spatial resolution is limited to the grid size. 
 In the Lagrangian description, one can instead be more accurate to much smaller scales in high-density regions.}

Our SR learning task is to enhance the Lagrangian spatial resolution by a factor of 8 and the mass resolution by 512. 
In other words, the SR field upsamples the number of tracer particles by $512$ times compared to the LR input, allowing us to save the much higher computational cost of running an HR $N$-body simulation.

\subsection{Dataset}

To train and validate our SR model, we prepare training and test sets with dark-matter-only $N$-body simulations using \texttt{MP-Gadget}\footnote{\url{https://github.com/MP-Gadget/MP-Gadget}}.
The $N$-body code solves the gravitational force with a split Tree-PM approach, where the long-range forces are computed using a particle-mesh method and the short-range forces are obtained with a hierarchical octree algorithm. 

\yueying{Our training and validation sets contain 16 and 1 LR-HR pairs of simulations respectively.}
The box size of all the simulations are $(100\,\hmpc)^3$, with $64^3$ LR and $512^3$ HR particles respectively. 
Here the LR is used as the input of SR model, and HR is used to train the discriminator by serving as one authentic realization of the high resolution field which shares the same large scale feature with their LR counterpart.
The mass resolution is $m_{\mathrm{DM}} = 2.98 \times 10^{11} \msun/h$ for LR, and $m_{\mathrm{DM}} = 5.8 \times 10^{8} \msun/h$ for HR, $1/512$ of the LR mass resolution.
We use $1/30$ of the mean spatial separation of the dark matter particles as the gravitational softening length.
The simulations have the WMAP9 cosmology \citep{hinshaw13} with matter density $\Omega _{\rm m} = 0.2814$, dark energy density $\Omega _{\Lambda} = 0.7186$, baryon density $\Omega _{\rm b} = 0.0464$, power spectrum normalization $\sigma_{8} = 0.82$, spectral index $n_{s} = 0.971$, and Hubble parameter $h = 0.697$. 
We train our model separately on snapshots at $z = 2$ and $z = 0$, to validate it at different levels of nonlinearity.

As for the test set, we ran another 10 different pairs of LR-HR simulations, of the same box size and cosmology as the training set.
Throughout this work, we test the performance of our trained models on these 10 test realizations in statistical comparison between the HR and SR results.

\subsection{Training and models}

The left panel of Figure~\ref{fig:training} shows the schematic of our training process.
We first preprocess the LR (HR) simulation by converting the particle positions to displacements,
which are then concatenated with the velocities of the tracer particles to form the input (output) as 3D fields with 6 channels.
The displacement and velocity fields are labelled by the Lagrangian particle positions on the grid $\mathbf{q}_i$. 
Due to the limitations of GPU memory, we crop the $100\, \hmpc$ simulation boxes into cubical chunks, of side length $\sim 20\, \hmpc$,
and pad extra margins around the LR input in a periodic fashion.
The latter compensates for the loss of edge voxels by the convolution operations, and preserves
translational symmetry.
The grid sizes of the fields are noted in the network architecture
diagram in Figure~\ref{fig:training}.

In our GAN model, the generator $G$ transforms an LR input $l$ to SR displacements and velocities $G(l)$ at $512\times$ the LR resolution,
while the discriminator $D$ evaluates the authenticity of
both the generated SR and the simulated HR realizations.
To this end, we use the Wasserstein GAN (WGAN) in which the loss function
is the distance between the distributions of real and fake images by optimal transport, known as the Wasserstein distance.
In WGAN, the discriminator is constrained to be Lipschitz continuous with a Lipschitz constant of 1.
WGAN is empirically superior to the vanilla GAN, as it is more stable, requires less tuning, \yueying{and has an informative loss function. A lower WGAN loss can in principle indicate a better model as the distance between the generated field distribution and the authentic distribution is shorter.}
But to maintain the Lipschitz constraint it requires more computation per batch.
The most popular variant adds a gradient penalty regularization
term in the loss function \citep[WGAN-gp;][]{wgan_gp}.
We train our SR networks using the WGAN-gp method, and for efficiency only penalize the
critic gradient every 16 batches.

Specifically, the WGAN-gp loss function is
\begin{multline}
  L_\mathrm{WGAN-gp} = \mathrm{E}_{l, z} [D(l, G(l, z))]
  - \mathrm{E}_{l,h} [D(l, h)] \\
  + \lambda \; \mathrm{E}_{l, h} \bigl[\bigl( \|\nabla_i D(l, i)\|_2 - 1 \bigr)^2\bigr].
  \label{L_GAN}
\end{multline}
The first line is the Wasserstein distance, and the second gives the gradient penalty, for which a random sample $i$ is drawn uniformly from the line segment between pairs of real ($h$) and fake ($G(l, z)$) samples,
\yueying{with the latter generated from LR samples $l$ and white noise maps $z$ (described in Sec.~\ref{sub:arch}).
$\lambda$ is a hyperparameter to balance the adversarial loss and the Lipschitz constraint, and we set it to the recommended and typical value 10.}
See \cite{wgan_gp} for more details on WGAN-gp.
Note that training $D$ involves minimizing all three terms of $L_\mathrm{WGAN-gp}$, whereas training $G$ only maximize the first one.

Since the HR and LR fields correlate on large scales, we can teach the discriminator the dependence of short modes on the long ones by feeding the LR field as additional input.
\yueying{As shown in Equation~\ref{L_GAN}, we condition $D$ on the LR input $l$ and make the networks a conditional GAN (cGAN). This helps $G$ to generate HR samples with right long- and short-wavelength mode coupling, i.e.\ forming smaller structures depending on the surrounding environments.}
In practice, we tri-linearly interpolate the LR field to match the size of the SR and HR fields before concatenating the upsampled field to both, respectively.

In addition, we also concatenate the Eulerian space density field to the discriminator,
computed from the displacement field using a differentiable Cloud-in-Cell operation.
The discriminator is then able to see directly structures in the Eulerian picture.
This is crucial for generating visually sharp features and predicting accurate small-scale statistics.
In order to concatenate a higher resolution Eulerian density field that resolves a scale smaller than the fundamental grid size ($L_{\mathrm{box}}/N_{\mathrm g}$),
we assign particles to a grid 2 times finer than the Lagrangian particle grid, use an ``inverse pixel shuffle''\citep{shi2016real} to re-shape the finer pixel values as extra channels, and concatenate the shuffled high-resolution density field to the input of the discriminator.
In this case, the discriminator takes in 20 channels in total (see the discriminator block in Figure~\ref{fig:training}), with 6+6 channels from the LR+SR $\{ \mathbf{d}_i, \mathbf{v}_i \}$ field and $8=2^3$ channels from the Eulerian density field.
We find this training scheme giving us the current best model.

\subsection{Details of the architecture}
\label{sub:arch}

The architecture of our GAN model follows those used in \cite{Li2021}.
We have released our framework \texttt{map2map}\footnote{\url{https://github.com/eelregit/map2map}} to train the SR model.
The structure of the generator network and its components is shown in the right two panels of Figure~\ref{fig:training}.
The colored plates represent different operations, connected by arrowed lines from the input to the output.
The sizes (channel number $\times$ spatial size) of the input, intermediate, and output fields are labelled.
The generator has a ladder shaped structure, with each rung upsampling the fields by a factor of two.
The left rail has 3 convolution blocks (blue ``conv'' plates),
operating in high-dimensional latent space.
The horizontal rungs project (yellow ``proj'' plates) the latent-space fields to the
low-dimensional output space on the right rail.
The projected results are then tri-linearly interpolated
(gray ``interp'' plates), before being summed together to form the output.
A crucial ingredient is the injection of white noise (red plates), that adds
stochasticity that are absent from the input.
These white noises are then correlated and transformed on different scales by
the subsequent convolution layers and activation layers.
The convolution kernel sizes are labeled in their plates, e.g., ``conv $3^3$''
for a $3\times3\times3$ convolution.
Throughout we use Leaky ReLU
with slope $0.2$ for negative values for the activation (green ``act'' plates).
All the ``conv'' blocks have the same structure as shown in the ``conv'' plate, except
the first one, which begins with an additional $1^3$ convolution (to transform the number
of channels) and an activation.
    
The discriminator network architecture is also illustrated in Figure~\ref{fig:training}.
It follows a fully convolutional ResNet-type structure \citep{he2016deep}.
Our residual block has two branches: the so-called ``skip''
branch is a $1^3$ convolution, and the ``residual''
branch consists of $3^3$ convolutions and activations.
The two branches are summed and then downsampled by $2\times$ with linear interpolation.
Other than the residual blocks, the first ``conv'' block is a
$1^3$ convolution followed by an activation, and the last ``conv''
block has a $1^3$ convolution to double the channel, an activation,
and a $1^3$ convolution to reduce to single channel.
Each output voxel values represent a score whose receptive field is a patch of the input to the discriminator.
Thus each score only evaluates part of the input field.
Class of fully convolutional discriminator like this is known as patchGAN \citep{isola2017image}.
Scores of all voxels are averaged to evaluate the Wasserstein distance in Equation~\ref{L_GAN}.



\section{Results}
\label{section3:Result}

\begin{figure*}
\centering
  \includegraphics[width=0.98\textwidth]{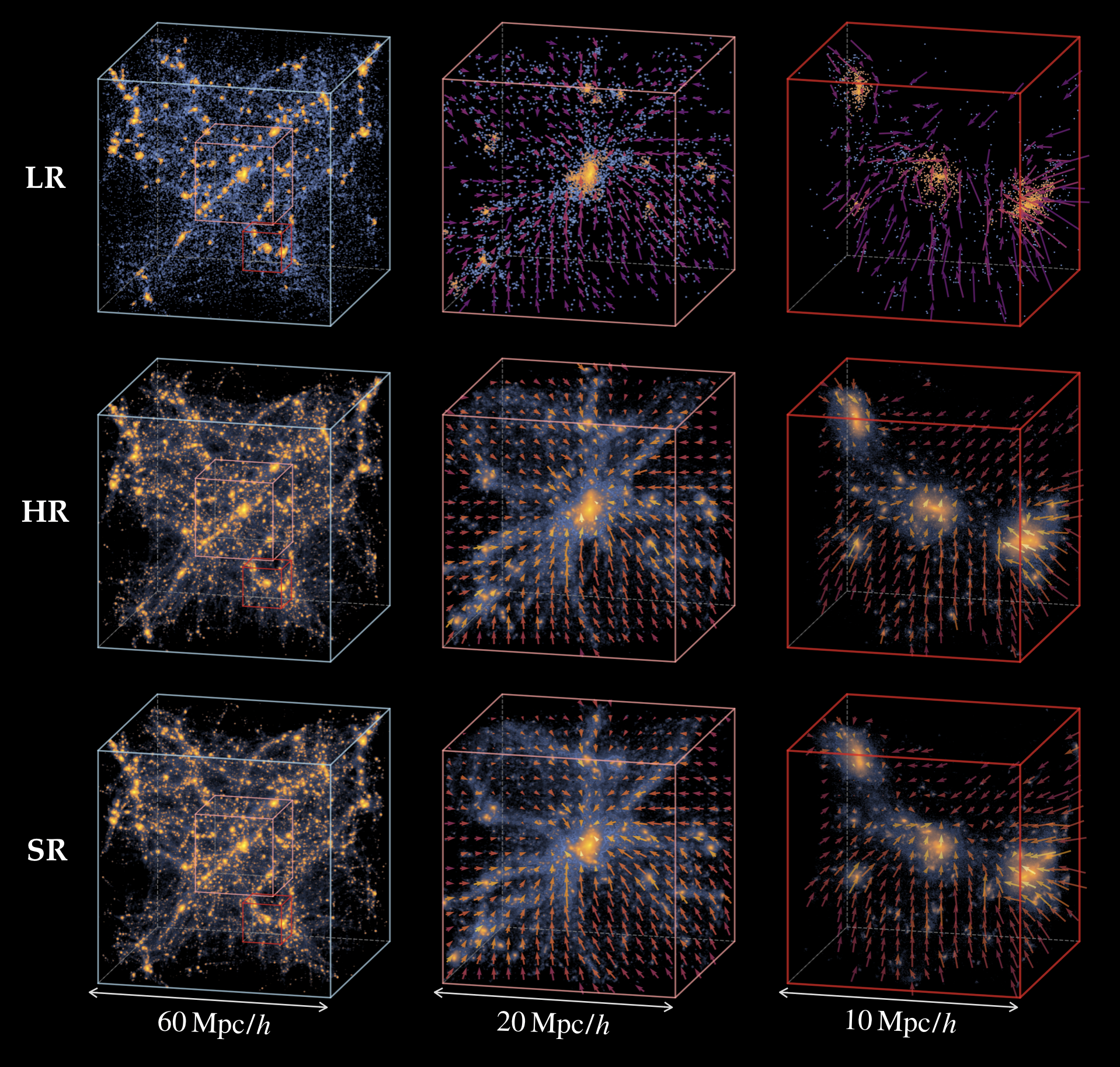}
  \caption{
    3D visualization of  low-, high-, and super-resolution (LR, HR, and SR) dark matter density and velocity fields at $z=0$.
    The left panels show a $(60\,\hmpc)^3$ sub-volume from one of the 100 $\hmpc$ test simulations to illustrate the density field on large scale.
    The blue background shows the density field of all the dark matter particles,
    smoothed with a Gaussian filter of width $5\, \hkpc$. 
    On top of this the particles in FOF groups are coloured orange to help visually identify the halos.
    The middle and right panels zoom into the pink and red boxes shown in the left panel, with sizes of $(20\,\hmpc)^3$ and $(10\,\hmpc)^3$, respectively, to reveal finer details of the structures and also to illustrate the velocity field.
    The arrows in the right two columns give the velocity field calculated by averaging the particle velocity in each voxel and projecting onto the image plane. 
    The color of the arrows is scaled by the particle number in that voxel, from purple to yellow indicating small to large particle numbers. 
    We only show velocity arrows for voxels with more than 200 particles for the HR and SR fields, and with more than 5 particles for the LR field.
    The top two rows show the LR and HR simulations, which share the same seed for initial conditions but are $512$ times different in mass resolution.
    The bottom panels show the SR realization generated by our trained model.
  }
  \label{fig:image-z0}
\end{figure*}

\begin{figure}
\centering
  \includegraphics[width=1.0\columnwidth]{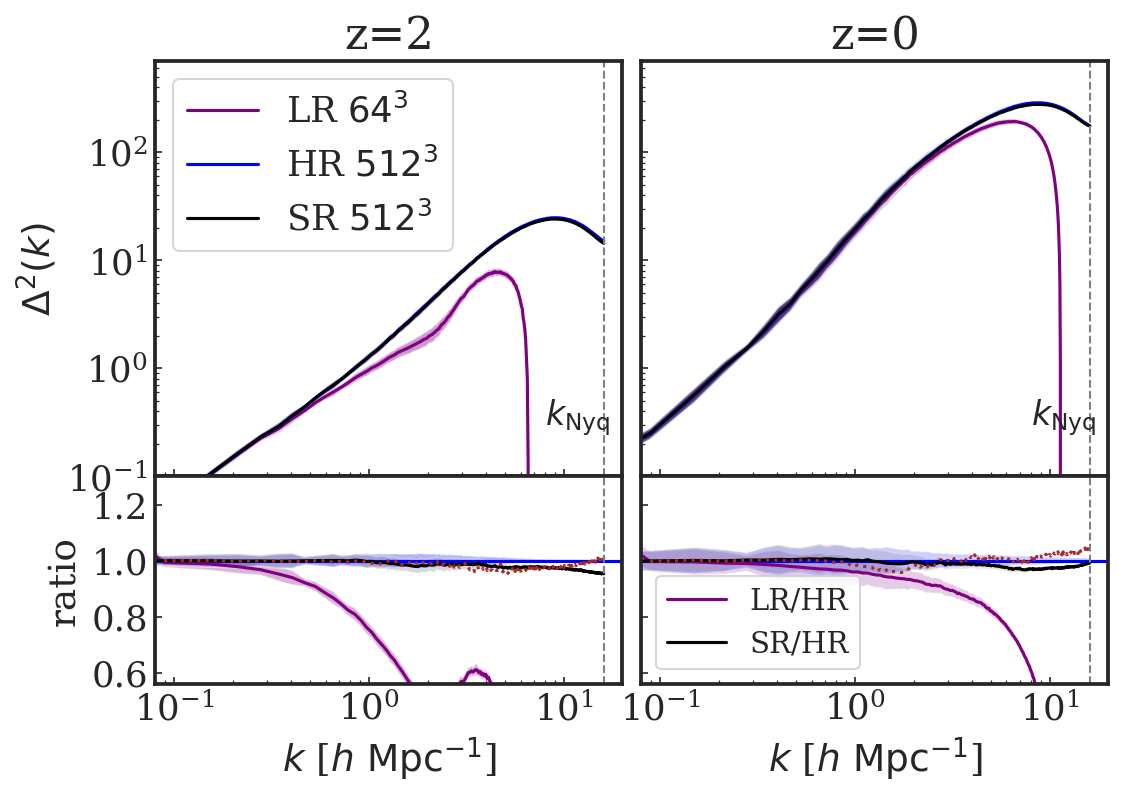}
  \caption{The upper panels give the dimensionless matter power spectrum $\Delta^2$ computed for the LR (purple), HR (blue) and SR (black) density field at $z=2$ and $z=0$.
  The vertical dashed lines mark the Nyquist wavenumber $k_{\mathrm{Nyq}} = \pi N_{\mathrm{mesh}}/L_{\mathrm{box}}$. The lower panels give the ratio between the SR and HR power spectrum. The SR power spectra match the HR curves within 5\% at both redshifts, giving a dramatic improvement compared with LR. 
  The shaded area shows the $1\sigma$ deviation measured from our 10 test sets.
  The brown dotted line in the lower panel is the result from our previous model in Paper I.}
  \label{fig:powerspectrum}
\end{figure}

\begin{figure*}
\centering
  \includegraphics[width=0.48\textwidth]{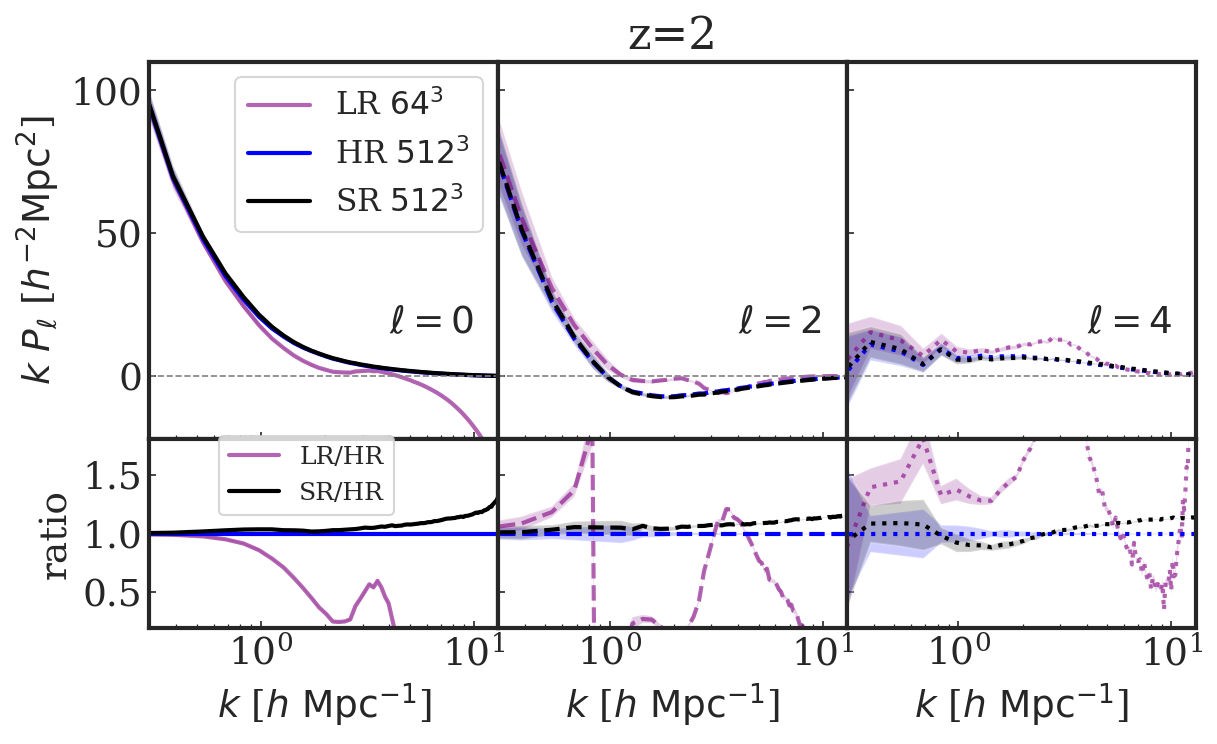}
  \includegraphics[width=0.49\textwidth]{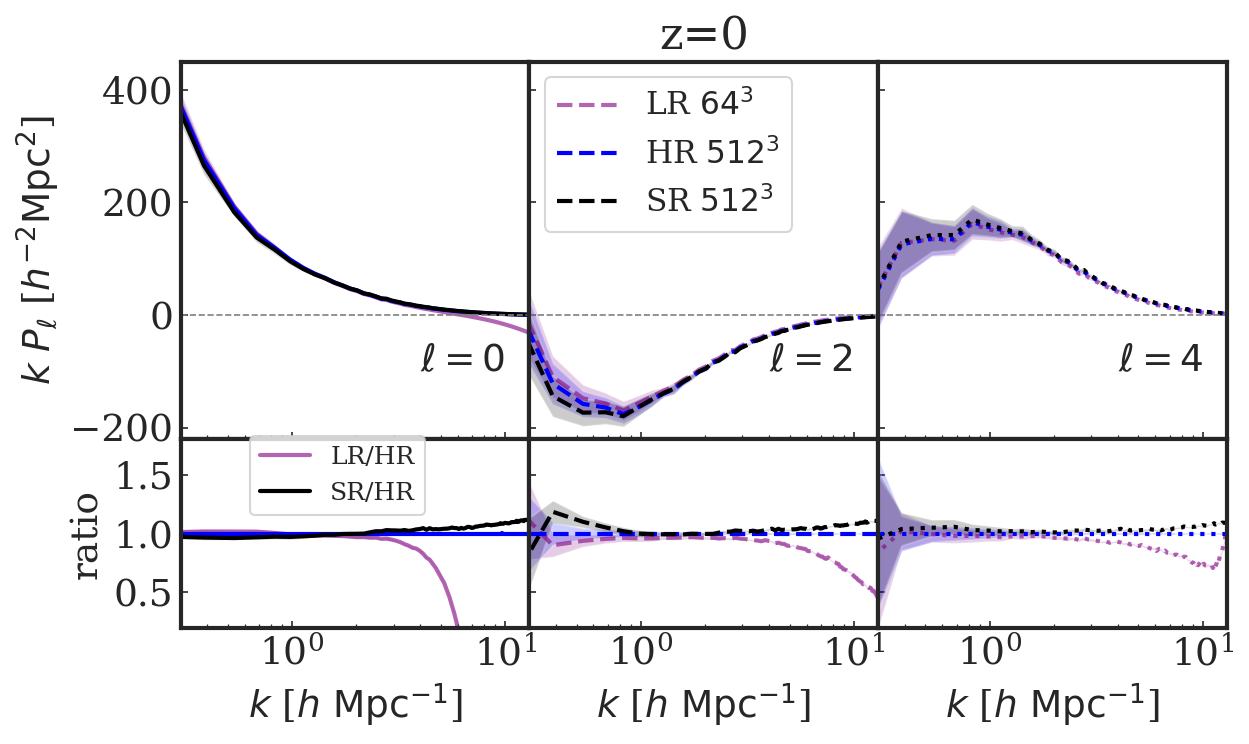}
  \caption{Multipoles of the 2D matter power spectrum with redshift space distortion along a specified line of sight, for LR (purple), HR (blue) and SR (black) fields at $z=2$ and $z=0$. $P_{\ell}$ projects the 2D power $P(k, \mu)$ onto a basis defined by Legendre weights, with $\ell = 0$ (solid line), $\ell = 2$ (dashed line) and $\ell = 4$ (dotted line) representing the order of the Legendre polynomial. The shaded area shows the $1 \sigma$ deviation measured from the 10 test sets.}
  \label{fig:rsd-ps}
\end{figure*}

\begin{figure*}
\centering
  \includegraphics[width=0.49\textwidth]{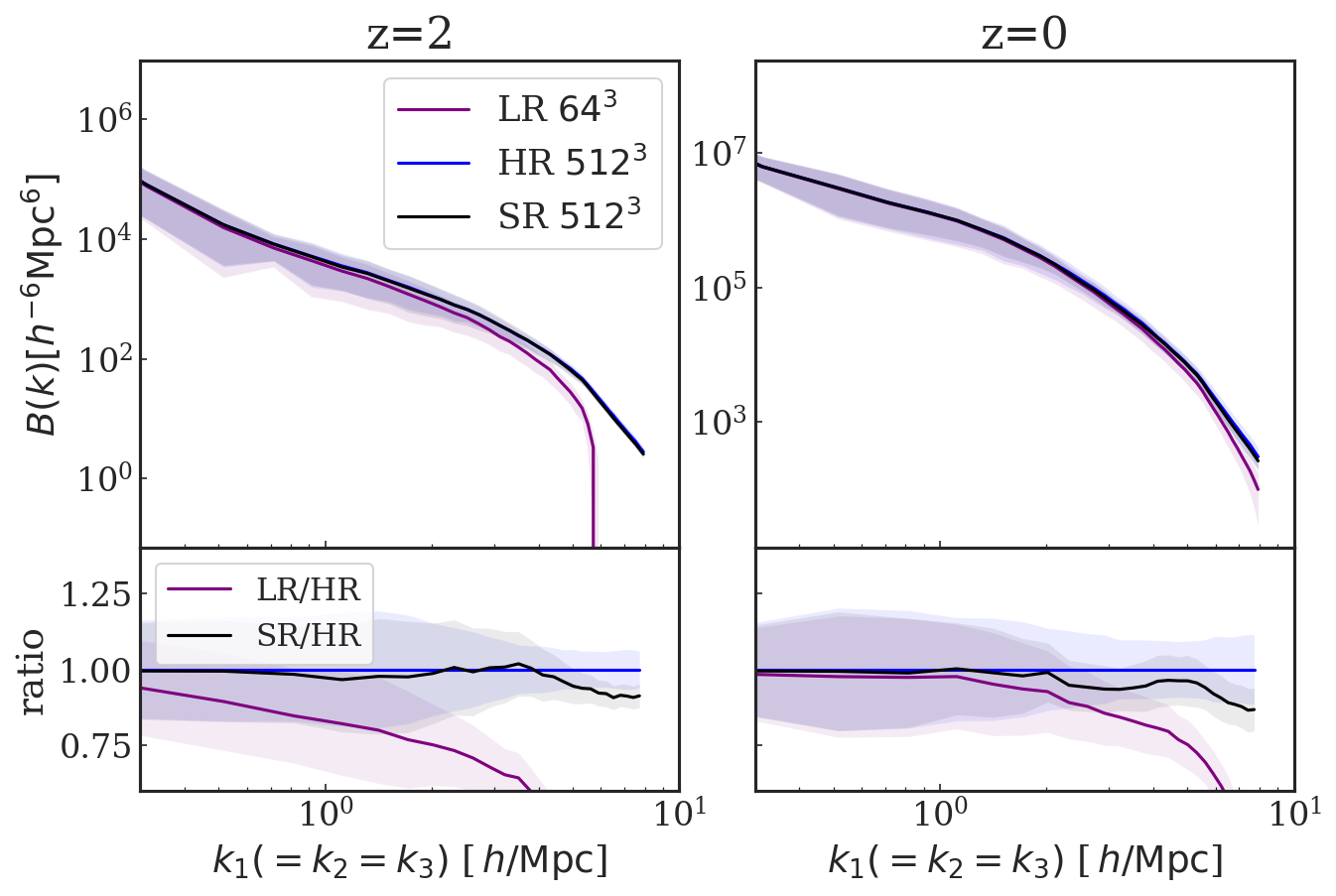}
  \includegraphics[width=0.49\textwidth]{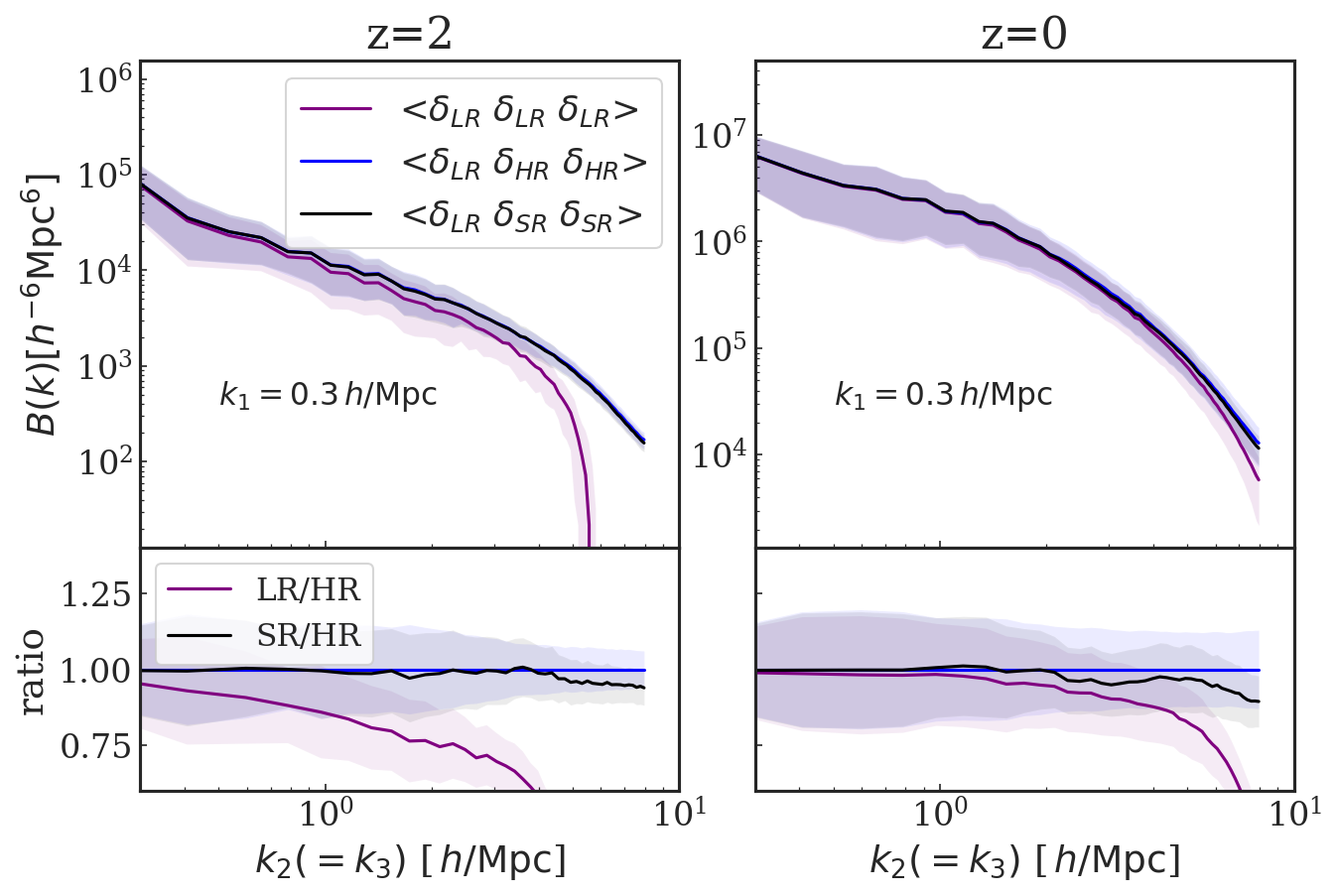}
  \caption{The statistics of the 3D bispectra for LR (purple), HR (blue) and SR (black) of test sets at $z = 2$ and $z = 0$. The shaded area gives $1 \sigma$ deviation measured from the test sets. 
  \textit{\textbf{Left panel:}} The bispectra given in equilateral configuration with $k_1 = k_2 = k_3$. 
  \textit{\textbf{Right panel:}} The cross bispectra between LR field and HR (the blue line), and between LR and SR field (the orange line) for the test sets. Here the bispectra are given in isosceles triangle configuration with fixed $k_1 = 0.3 \hmpc$ and $k_2 = k_3$.}
  \label{fig:Bispectrum}
\end{figure*}

\subsection{Visual comparison}
\label{subsection:visual}

\yueying{As a first validation, we visually compare one generated SR field to an authentic HR realization from an $N$-body simulation.}
Figure~\ref{fig:image-z0} shows 3D visualizations of the dark matter density and velocity field from one test realization at $z = 0$. 
The images are rendered with \texttt{gaepsi2}\footnote{\url{https://github.com/rainwoodman/gaepsi2}}.
The top panels show the LR simulation used as the input to generate the SR realization shown in the bottom ones. 
The middle row shows an HR realization, from an HR simulation that has $512\times$ higher mass resolution but shares the same random seed (same modes at small $\mathbf{k}$ or on large scales) as LR in the initial conditions. 

The left column of Figure~\ref{fig:image-z0} shows the LR, HR, and SR density fields in a $(60\,\hmpc)^3$ sub-box.
We identify halos with the Friends-of-Friends (FOF) halo finder and highlight the halos in orange.
To illustrate finer details of the evolved structure and also to visualize the velocity field, in the right two columns we zoom into two smaller sub-boxes of size $(20\,\hmpc)^3$ and $(10\,\hmpc)^3$, centered on two massive halos, as shown by the pink and red boxes in the left panel, respectively.
The colored arrows show the spatially averaged velocity field, calculated by evenly dividing the box into voxels and projecting the averaged particle velocities in the voxels onto the image plane. 
The length of the arrows is scaled by the projected velocity magnitude, and the color by the number of particles in the corresponding voxel, from purple to yellow indicating small to large particle numbers. 
The velocity field shown in the two sub-volumes coherently points to the central large halo, as the matter field is falling into the large structure under gravity. 

The visualization of the density field contains morphological and high-order statistical information
that the human eyes can easily recognize and classify.
This visual comparison between the HR and SR fields shows that our model can generate visually authentic small scale features conditioned on the large-scale environment set by the LR input.
The SR field is able to form halos where the input LR can not resolve them, and thus enables us to probe a halo mass range enhanced by orders of magnitude, making our method promising for a variety of applications based on halo (and subhalos) catalogs, as we will further quantify in the following sections.


\subsection{Full field statistics}
\label{subsection:full-field}

\subsubsection{Matter power spectrum}

The matter power spectrum is one of the most commonly used summary statistics for the matter density field.
It is the Fourier transform of the 2-point correlation function $\xi(r)$ defined as
\begin{equation}
\begin{split}
    \xi(|\mathbf{r}|) &= \left\langle\delta\left(\mathbf{r}^{\prime}\right) \delta\left(\mathbf{r}^{\prime}+\mathbf{r}\right)\right\rangle \\
    P(|\mathbf{k}|) &= \int \xi(\mathbf{r}) \mathrm{e}^{i \mathbf{k} \cdot \mathbf{r}} \mathrm{d}^{3} \mathbf{r}
\end{split}
\end{equation}
where $\delta(\mathbf{r}) = \rho(\mathbf{r})/\bar{\rho} -1$ is the density contrast field, $\mathbf{k}$ is the 3D wavevector of the plane wave, and its  magnitude $k$ (the wavenumber) is related to the wavelength $\lambda$ by $k = 2\pi/\lambda$.
$P(k)$ quantifies the amplitude of density fluctuations as a function of scale, and that encodes the complete summary information for a Gaussian random field. 
$P(k)$ can be used as a reliable metric to evaluate the fidelity of the SR field.

In Figure~\ref{fig:powerspectrum}, we compare the dimensionless matter power spectra $\Delta^2 (k) \equiv k^3 P(k)/2\pi^2$ at $z=2$ and $z=0$.
The vertical dashed lines mark the Nyquist wavenumber $k_{\mathrm{Nyq}} = \pi N_{\mathrm{mesh}}/L_{\mathrm{box}}$ with $N_{\mathrm{mesh}}=512$ and $L_{\mathrm{box}} = 100\,\hmpc$.
The monotonically increasing $\Delta^2(k)$ is equivalent to the contribution (in $\log(k)$ bins) of different scales to the variance of the matter density fluctuation, and thus a useful indicator that divides the linear and nonlinear (towards increasing $k$) regimes at $k_\mathrm{NL}$ where $\Delta^2(k_\mathrm{NL}) \sim 1$. 

We compute the density field by assigning the particle mass to a
$512^3$ mesh using the CIC (Cloud-in-Cell) scheme, for each of the LR, HR, and SR fields.
Note that here we do not apply the deconvolution correction for the resampling window of mass assignment \cite{Jing05}, as that would amplify the noise and leads to artifacts in the LR results.

The SR power spectra successfully matches the HR results to percent level (differed by less than 5\%) to $k_\mathrm{max}\approx16\,\hmpc$ at all redshifts, a dramatic improvement over the LR predictions. This is remarkable considering the fact that the SR model fares equally well from the linear to the deeply nonlinear regimes.
\yueying{We compare the matter power of the HR and SR fields down to a scale of $k_\mathrm{max} \sim 16 \hmpc$. This corresponds to the Nyquist frequency of the HR (SR) simulation and that beyond that point there is not justification to require agreement.}
The brown dashed line in Figure~\ref{fig:powerspectrum} shows the test set power spectrum from the model of \cite{Li2021} which only generates the displacement field of tracer particles. 
The model presented in this work produces the full phase space of tracer particles while yielding a generated matter density field of equivalently fidelity (measured using P(k)) to our previous model.

\yueying{
The shaded areas in the bottom panels of Figure~\ref{fig:powerspectrum} show the standard deviation of the ratio compared to the averaged HR power spectrum.
The average value and its deviation are measured over the 10 test sets.
We can see that the variance of the amplitude of power coming from the HR test sets is comparable with that from the SR sets, implying a reasonable 4-point function between the HR and SR simulations.}
\yueying{Beyonds this,} we also find that SR variants (using different random input noise) have almost identical power (within $\sim 1 \%$).
This is because sample variance on small scales is suppressed under the large-scale constraint.

\subsubsection{Redshift-space power spectrum}

Redshift space distortion (RSD) is an important probe of cosmology from ongoing and upcoming large scale structure surveys (see e.g., \citealt{percival11} for a review).
With the full phase space dynamic field generated by our SR model, we can examine the 2D power spectrum of the RSD matter field, which encodes the velocity field information as further validation of our generated SR field. 
Here we analyze the entire dark matter field in redshift space, and will further discuss the redshift space halo correlation in Section~\ref{subsubsection:halo-rsd}.

For an $N$-body simulation output, the distortion from real space to redshift space  is performed by moving the particles along one specific direction (line of sight) according to their peculiar velocity.
\begin{equation}
\label{equation:rsd}
    \mathbf{s} = \mathbf{x} + \frac{v_z}{aH(a)} \hat{z},
\end{equation}
where $\mathbf{s}$ is the redshift-space coordinate, $\mathbf{x}$ the
real-space counterpart, and $v_z$ is the peculiar velocity of the tracer particles, whereas $\hat{z}$ denotes the unit vector along the line of sight.

We measure the 2D matter power spectrum $P(k,\mu)$ of the RSD matter density field.
Here $\mu$ is defined with respect to the line of sight $\mu = k_{\parallel}/k$ and $k_{\parallel}$ is the component of the wavevector $k$ along the line of sight. We split $\mu$ into 5 bins ranging from $\mu = 0$ to $\mu = 1$.

Figure~\ref{fig:rsd-ps} gives the resultant 2D power spectrum by projecting $P(k,\mu)$ onto the basis of Legendre Polynomials
\begin{equation}
    P_{\ell}(k)=(2 \ell+1) \int_{0}^{1} d \mu P(k, \mu) \mathcal{L}_{\ell}(\mu)
\end{equation}
where $\mathcal{L}_{\ell}$ is the order of Legendre Polynomials represented by different line styles.
The lines in each panel show the mean value measured from the 10 test sets and the shaded area is the $1\sigma$ standard deviation.

$P_{\ell}$ for the LR sets drops off quickly when extending to large $k$ due to the lack of small scale 
fluctuations. 
\yueying{The monopole moment $P_{\ell=0}$ of LR goes negative on  small scales because of the subtraction of the shot noise.}
Meanwhile the SR field predicts a $P_{\ell}$ that matched the HR results well, with only a small excess of power one the smallest scale. At $k = 10 \invhmpc$, the averaged SR $P_{\ell}$ prediction is about 13\% higher than HR at $z=2$ and 8\% higher than at $z=0$.

\subsubsection{Bispectra}

As they are products of non-linear structure evolution, we need higher order statistics to characterize the non-Gaussian features of the evolved density fields.
One of the fiducial tools used to probe higher order statistics and to quantify the spatial distribution of the cosmic web is the bispectrum \citep{Scoccimarro2000,Bernardeau2002}.
It is the Fourier transform of the three point correlation function, defined as
\begin{equation}
    (2 \pi)^{3} B\left(\boldsymbol{k}_{1}, \boldsymbol{k}_{2}, \boldsymbol{k}_{3}\right) \delta_{\mathrm{D}}\left(\boldsymbol{k}_{1}+\boldsymbol{k}_{2}+\boldsymbol{k}_{3}\right)=\left\langle\delta\left(\boldsymbol{k}_{1}\right) \delta\left(\boldsymbol{k}_{2}\right) \delta\left(\boldsymbol{k}_{3}\right)\right\rangle
\end{equation}
where $\delta_D$ is the Dirac delta.

In Figure~\ref{fig:Bispectrum}, we compare the bispectra for the LR, HR and SR density fields at $z=2$ and $z=0$ separately.
The solid lines in each panel shows the mean value measured from the 10 test sets while the shaded area shows the corresponding $1\sigma$ standard deviation.

The left panel of Figure~\ref{fig:Bispectrum} shows bispectra for an equilateral configuration, $k_1 = k_2 = k_3$. 
$B(k)$ for LR drops off quickly when extending to large $k$, as the LR field lacks fluctuations on small scales and $B(k)$ is cut off by the large shot noise in LR field.
On the other hand, the SR field gives good agreement with the HR result. At a small scale of $k \sim 8 \invhmpc$, the mean $B(k)$ for SR field differs from the HR field by about 7\%. 
We  note that the difference between the HR and SR field for different realizations
exhibits relatively large variations ($\pm 10$ \%) on small scales. These differences are an order of magnitude larger than those for the power spectrum of different realizations (Figure \ref{fig:powerspectrum}). This indicates that the small scale fluctuations that differ between realizations are more sensitively probed by higher order statistics.

In the right panels of Figure~\ref{fig:Bispectrum}, we examine the cross bispectra between the LR and HR fields (blue line), and between LR and SR fields (black line) for the test sets at $z = 2$ and $z = 0$. 
The bispectra is given for an isosceles triangle configuration, with fixed $k_1 = 0.3 \hmpc$ and $k_2 = k_3$. 
The notation used, $\langle \delta_{\mathrm{LR}}$ $\delta_{\mathrm{HR}}$ $\delta_{\mathrm{HR}} \rangle$ indicates that $k_1$ is sampled from the LR field while $k_2$, $k_3$ are from the HR field. 
For comparison, we also show $\langle \delta_{\mathrm{LR}}$ $\delta_{\mathrm{LR}}$ $\delta_{\mathrm{LR}} \rangle$ as a purple line.

The cross bispectrum quantifies the response of the modes of the HR (SR) field to those of the LR field.
With the isosceles triangle configuration, $k_2 = k_3 >> k_1$ corresponds to the squeezed limit that characterizes the response of the small scale modes to the large scale modes.
For the cross bispectra, the SR field again matches the HR results well, within the $1 \sigma$ standard deviation. 
At the squeezed limit of $k \sim 8 \invhmpc$, the mean $B(k)$ for the SR field differs from the HR field by about 6\%. 
The good agreement between HR and SR implies that our generated SR field can reproduce the high order statistics of the density field with fidelity.

\subsection{Mock halo catalog analysis}
\label{subsection: halo-catalog}

\begin{figure*}
\centering
  \includegraphics[width=1.0\textwidth]{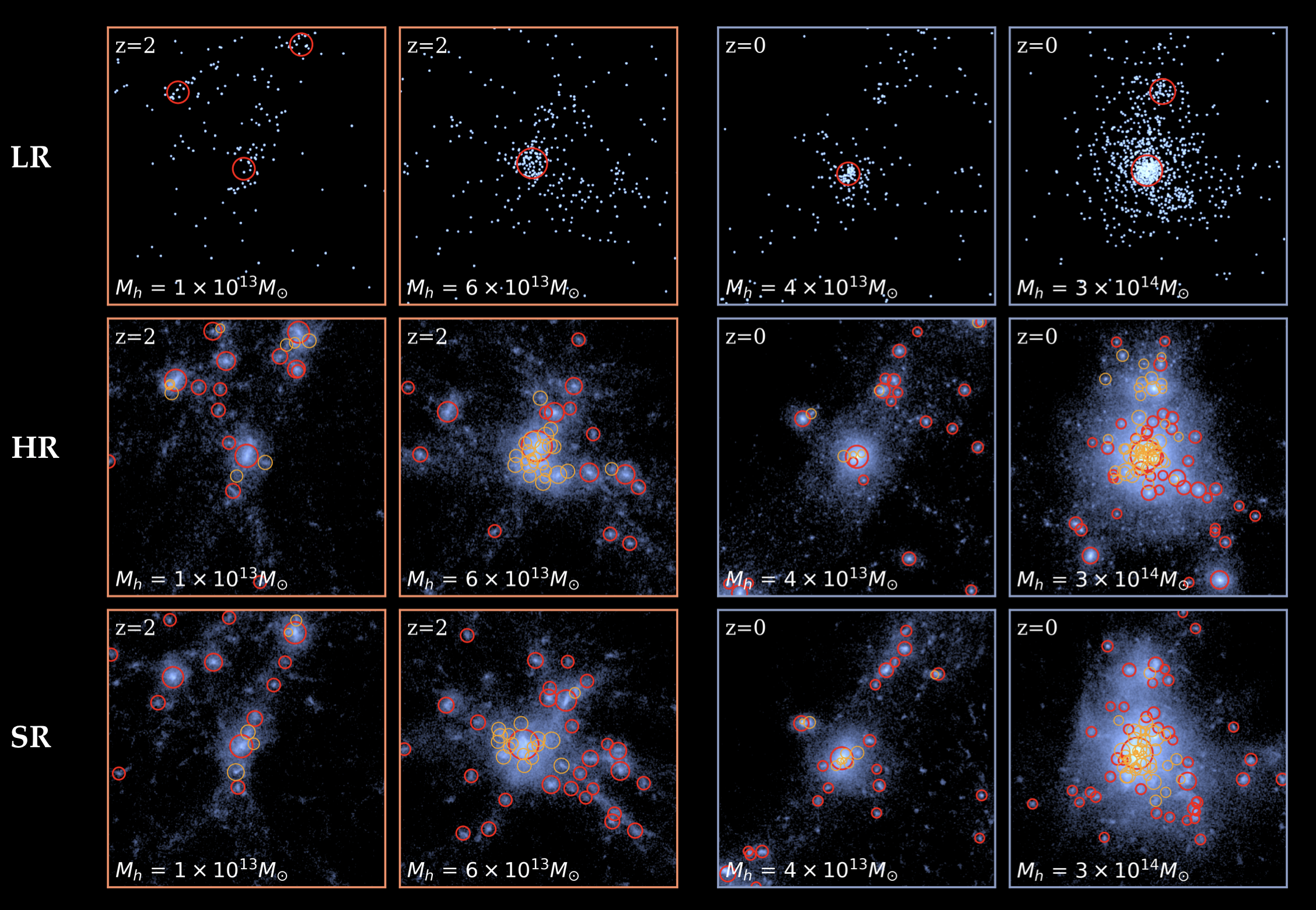}
  \caption{Visualization of the small scale substructures of the LR (top panel), HR (middle panel) and SR (bottom panel) fields in the test set at $z=2$ and $z=0$. All the panels shown here are $(5\,\hmpc)^3$ in size centered on a massive halo with the halo mass $M_{\mathrm{h}}$ given in the legend. The blue backgrounds show the projections of the smoothed density fields rendered by \texttt{gaepsi2}, 
  We employ the AHF (see text) to identify halos and subhalos in each field and mark the host halos with red circles and subhalos with orange circles to help identify the substructures. The radius of the circle is proportional to the virial radius, $R_{\mathrm{vir}}$. All  halo and subhalos shown here have virial mass $M_{\mathrm{h}} > 2.5\times10^{11} \msun$, that corresponds to 300 tracer particles in HR (SR) field.}
  \label{fig:sub-halo-illustration}
\end{figure*}

\begin{figure*}
\centering
  \includegraphics[width=0.485\textwidth]{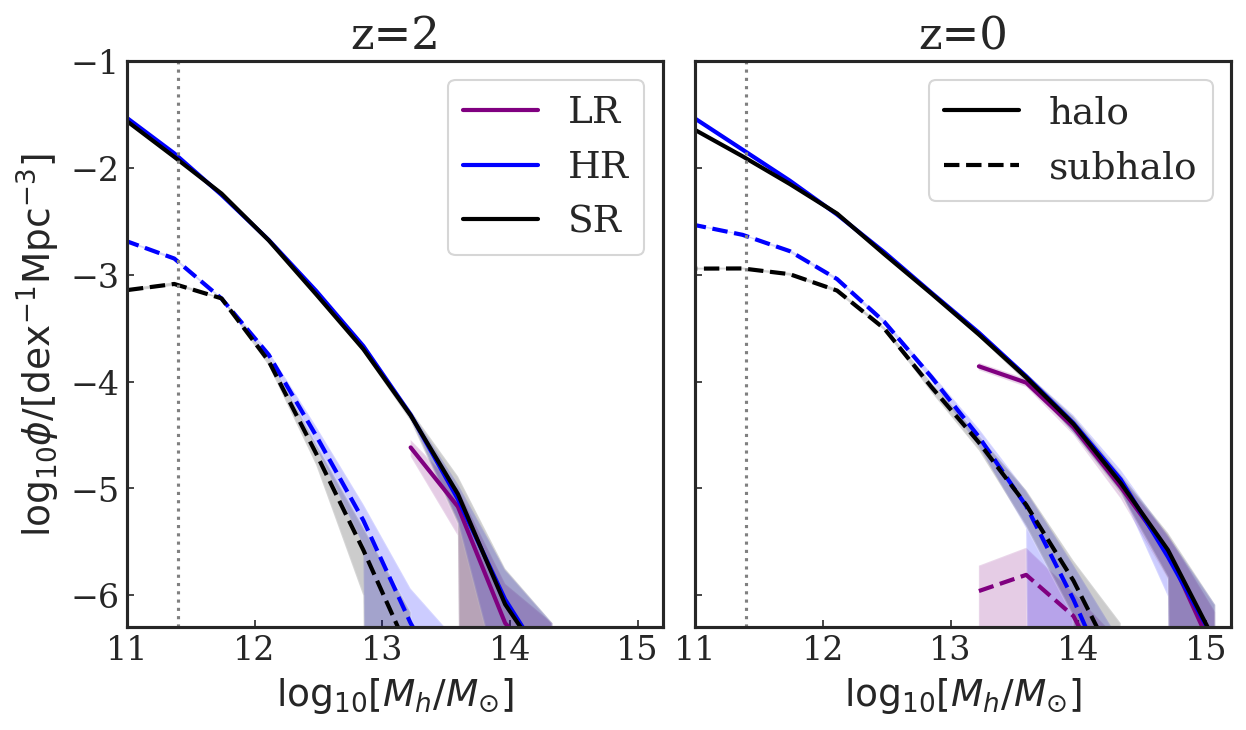}
  \includegraphics[width=0.495\textwidth]{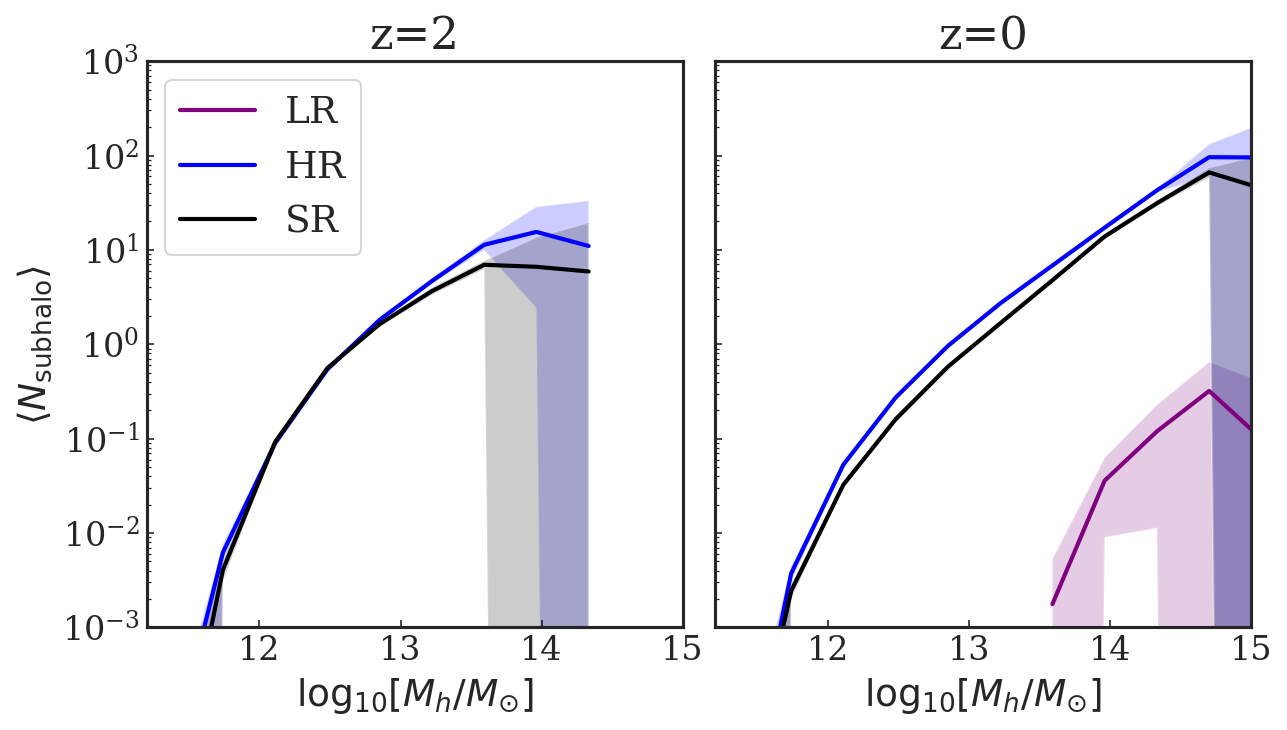}
  \caption{\textit{\textbf{Left panel:}} The halo and subhalo populations of the LR (purple line), HR (blue line) and SR (black line) field at $z=2$ and $z=0$. The solid lines show the populations of all the halos (including subhalos), and the dashed lines only include the subhalos. The vertical dotted line marks where $M_{\mathrm{h}} = 2.5 \times 10^{11} \msun$, corresponding to 300 tracer particles within our HR (SR) mass simulation. The shaded area shows the $1\sigma$
  standard deviation measured from the 10 test sets.
  \textit{\textbf{Right panel:}} 
  Mean occupation number of subhalos vs host halo mass the for LR, HR and SR fields at $z=2$ and $z=0$.
The  $y$ axis is the averaged number of subhalos with $M_{\mathrm{h}} > 2.5\times10^{11} \msun$ in each host halo mass bin. 
  The shaded area shows the $1\sigma$ standard deviation from the test sets.}
  \label{fig:sub-halo-mf-hod}
\end{figure*}

Dark matter halos contain a lot of substructure.   
Subhalos are the most important, corresponding to local maxima of the density field inside halos. The subhalos are expected to contain galaxy hosts and can be related to the observed galaxy population. 
In cosmological studies, mock galaxy catalogues are constructed (via various analytical or semi-analytical techniques) based on the dark matter halo and subhalo catalogs extracted from $N$-body simulations to compare to a variety of observables and to infer cosmological parameters from observed galaxy populations.

In this section, we generate the halo and subhalo catalogs from the 10 test sets of LR, HR and SR fields and examine the statistical properties of the halo and subhalo populations. Understanding how well substructure is generated is an important validation of our SR models.

We identify dark matter halos (and subhalos) in our LR, HR and SR fields using the \texttt{Amiga} halo finder (AHF) \citep{Knollman2009}. 
AHF uses adaptive refined meshes to identify local density peaks as centers of prospective halos (subhalos) and defines the halos as spherical overdensity regions that enclose density $\Delta_{\mathrm{vir}}$ times larger than the average background matter density:
\begin{equation}
    \frac{M(<R_{\mathrm{vir}})}{{\frac{4\pi}{3}}R_{\mathrm{vir}}^3} = \Delta_{\mathrm{vir}} \rho_{\mathrm{b}}
\end{equation}
where we use $\Delta_{\mathrm{vir}} = 200$, and $R_{\mathrm{vir}}$ is the virial radius of the halo, \yueying{and $\rho_\mathrm{b}$ is the average matter density of the universe.}
Throughout this work, we examine halo and subhalo populations with virial mass $M_{\mathrm{h}} \ge 2.5 \times 10^{11} \msun$ (i.e. which contain at least 300 tracer particles in our HR (SR) test set). 
This value corresponds roughly to the typical host mass of emission line galaxies \citep[ELGs, see e.g.,][]{rodrigues16}.
We specify this mass threshold as a sufficiently resolved halo in our HR (SR) set. 


\subsubsection{Visualization}

To visualize the substructure we identify,  Figure~\ref{fig:sub-halo-illustration} shows the matter density field for 
the LR, HR and SR fields within a $(5\,\hmpc)^3$ sub-volume centered around a massive halo. 
The panels in each column show the same region in the test sets of LR HR and SR,
We select 2 sub-regions around different halos at $z=2$ and $z=0$. One sub-region is for a halo from the most massive bin and the other is from a moderate mass bin. We use these two examples to illustrate the nature of substructures in different environments.
We mark all the host halos with red circles and the subhalos with orange circles, with the radius of the circle proportional to the virial radius of the halo $R_{\mathrm{vir}}$.

Comparing the LR, HR and SR fields in each column, we find that the LR field only forms the skeleton of the large scale structures while the poor resolution can barely capture any small halos or substructure.
However, conditioned on these large scales, the SR model is capable of generating small scale structures identified as gravitational bound objects. 
These appear visually as authentic when compared to the samples of the HR field obtained from a real $N$-body simulation.

We note again that, the HR field shown here is one true sample of a high resolution realization that shares the same large scale features as the LR field.
We emphasize that the task of the SR model is to generate realizations of high resolution field that {\it statistically} match the HR distribution instead of expecting one to one correspondence between HR and SR substructures that form below the LR resolution limit.
In the following subsection, we validate the generated SR field by examining different statistical properties of the halo substructure including its clustering, based on the halo catalogs derived from the SR and the HR fields.

\begin{figure}
\centering
  \includegraphics[width=1.0\columnwidth]{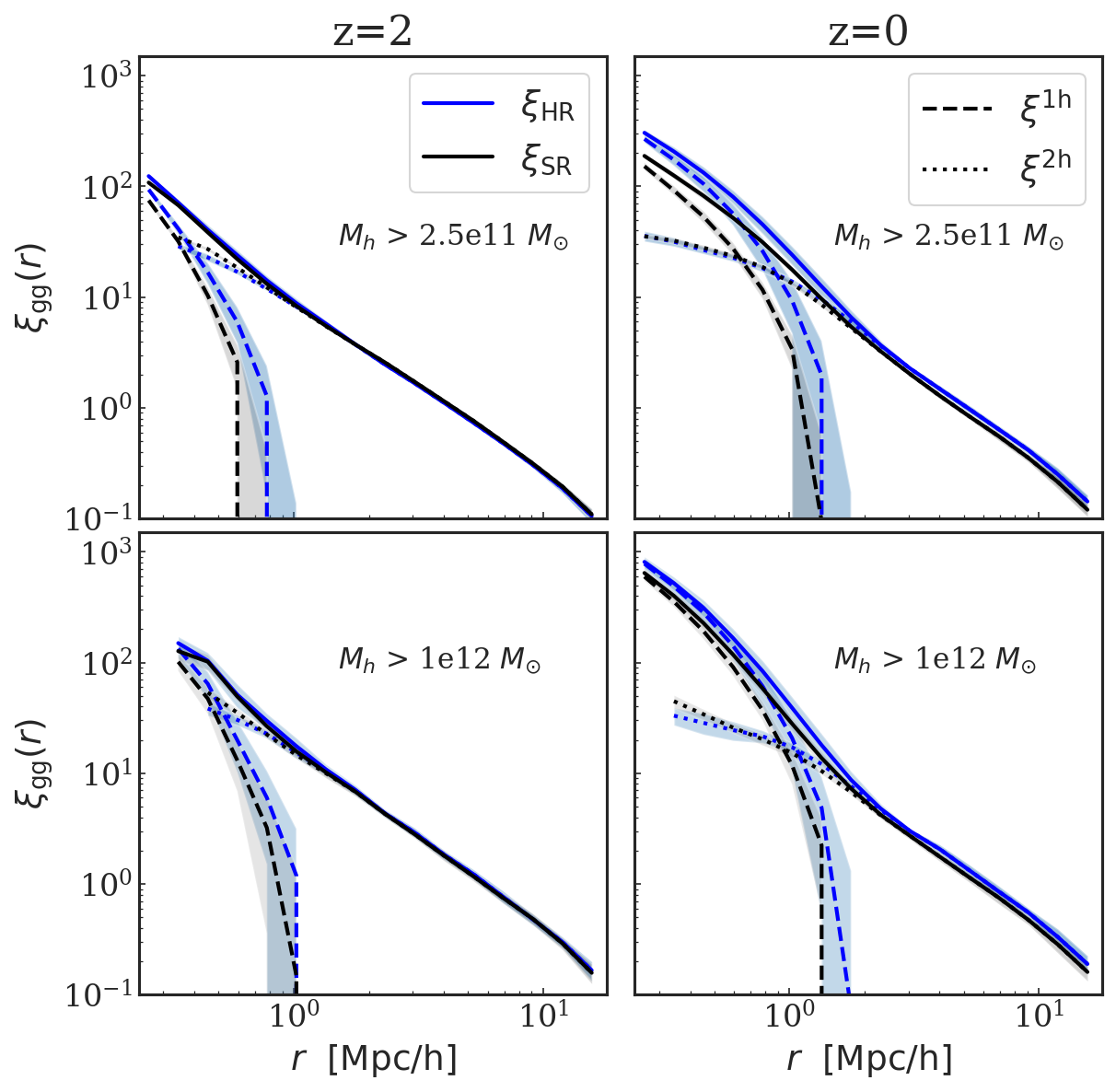}
  \caption{The two point correlation function of the halo distribution from HR (blue lines) and SR (black lines) field at $z=2$ and $z=0$. The values of $r$ in $x$ axis give the spatial distance between two objects in comoving coordinate.
  $\xi(r)$ is calculated based on the halo catalog identified by AHF. 
  The upper panels show $\xi(r)$ for all the halo (and subhalos) population with $M_{\mathrm h} > 2.5 \times 10^{11} \msun$  which corresponds to 300 dark matter particles, and the lower panels apply a higher $M_{\mathrm h}$ threshold of $10^{12} \msun$.
  In all the panels, the dashed lines are for the 1-halo component and the dotted lines for the 2-halo component. The shaded area shows the $1\sigma$ standard deviation measured from the 10 test sets.}
  \label{fig:Halo-Corr}
\end{figure}

\subsubsection{Halo and subhalo abundance}

Figure~\ref{fig:sub-halo-mf-hod} shows the abundance of  halos and subhalos as measured from the LR, HR, and SR $(100\,\hmpc)^3$ test set fields at $z=2$ and $z=0$.
The halo mass function is defined by $\phi \equiv N / \log_{10}\!M_\mathrm{h} / V_{\mathrm{box}}$, where $N$ is the number of halos above threshold mass $M_\mathrm{h}$ and $V_{\mathrm{box}}$ is the comoving volume of the simulation.
The purple, blue and black lines show the mean value from the LR, HR and SR test set results, respectively, with the shaded areas giving the $1\sigma$ standard deviation. 
The vertical dotted line corresponds to a halo mass of $M_{\mathrm{h}} = 2.5 \times 10^{11} \msun$, which we specify as the mass threshold for a resolved (300 particle) halo in our HR (SR) test set. 

The solid lines in the left two panels of Figure~\ref{fig:sub-halo-mf-hod} show the mass function of all the halos (including subhalos) identified by AHF.
Because of the large particle mass and low force resolution, the LR simulations only resolves halos above $M_{\rm h} \sim 10^{13} \msun$.

Using our GAN model, the SR field generated from the LR input is able to generate  halos over the whole mass range down to $M_{\mathrm h} 
\sim 10^{11} \msun$.
The generated SR fields predict overall halo populations that agree remarkably well with the HR results. Close to the low mass limit $M_{\mathrm{h}} = 2.5 \times 10^{11} \msun$, the SR fields have a slightly lower halo abundance than the HR fields (8\% lower at $z=2$ and about 13\% lower at $z=0$). This is mostly due to a deficit in the number of subhalos around this mass.

The left two panels of Figure~\ref{fig:sub-halo-mf-hod} show also the mass function of subhalos identified by the AHF in dashed lines.
Again the LR simulation cannot resolve subhalos properly due to the poor mass resolution. 
Our SR field predicts a subhalo abundance that statistically matches the HR field down to $M_{\rm h} > 10^{12} \msun$.
However at the lower mass end the SR model misses some of the subhalo population especially at $z=0$. 
For the overall abundance of subhalo within mass range of [$2.5 \times 10^{11} \msun$, $10^{12} \msun$], our SR field predicts subhalo abundance about 12\% lower at $z=2$ and about 40\% lower at $z=0$.
This is also evident in the illustrative example in Figure~\ref{fig:sub-halo-illustration}: the SR field has fewer subhalos (marked with orange circles) than the HR.

Figure~\ref{fig:sub-halo-mf-hod} shows the mean occupation number of subhalos as a function of the host halo mass.
Here the vertical axis is $\langle N_{\mathrm{subhalo}} | M_{\mathrm{host}} \rangle$, corresponding to the expected number of subhalos with $M_{\mathrm{h}} > 2.5\times10^{11} \msun$ in each host halo mass bin. 
The shaded area shows the $1\sigma$ standard deviation from the test sets.
At $z=2$, the occupation numbers are consistent. 
At $z=0$, we find an offset in $N_{\mathrm{subhalo}}$ for the SR field compared to HR.
For the host halo population in the mass range of [$10^{12} \msun$, $10^{15} \msun$], the mean number of subhalos in the SR field is about 20\% to $\sim$ 40\% below that of the HR field.

However, the deficit in subhalos and halo occupation number should not be fundamental,
but in principle fixable.
Recall that we feed the generator outputs to the discriminator network also in the Eulerian description.
The fact most missing halos are low-mass implies that the results can be improved
once our SR model can evaluate the Eulerian fields in higher resolution
and thus better resolves the small substructures.
We leave this for future improvement, with which statistically accurate mock catalogs can be released.



\subsubsection{3D halo correlation function}

Apart from the 1-point statistics of halo abundance, it is also important to quantify the spatial correlation function of the halos and their subhalos.
In Figure~\ref{fig:Halo-Corr}, We first measure the 3D spatial correlation function $\xi(r)$ for the halo population (with $M_{\rm h} > 2.5 \times 10^{11} \msun$) in the HR and SR field at $z=2$ and $z=0$.
Here we do not include the result from LR fields because the mass resolution is too poor (LR only has halos with $M_{\rm h} > 10^{13} \msun$) and does not have information on the spatial distribution of halos comparable to the HR simulation.
\yueying{We calculate the two point halo correlation $\xi(r)$ using the Landy-Szalay estimator \citep{Landy1993}.}
\yueying{
\begin{equation}
    \xi_{\text{LS}}(r) = \frac{DD(r)-2DR(r)+RR(r)}{RR(r)}
\end{equation}
}
where $\mathrm{DD}(r)$ is the number of sample (data) pairs with separations equal to $r$, $\mathrm{RR}(r)$ is the number of random pairs with the separations $r$, \yueying{and $\mathrm{DR}(r)$ measures the pair separation between the data set and the random set.}

$\xi(r)$ can be further decomposed into one-halo and two-halo term contributions respectively, with 
\begin{equation}
    \xi(r) = \xi^{1h} (r) + \xi^{2h} (r) 
\end{equation}
where $r$ is the comoving distance between halos, $\xi^{1h} (r)$ is the correlation function for subhalos which reside in the same halo, and $\xi^{2h} (r)$ accounts for those that reside in separate halos.
The dashed and dotted line in Figure~\ref{fig:Halo-Corr} give the $\xi^{1h} (r)$ and $\xi^{2h} (r)$ component separately.
For each redshift, we show the halo correlation function for the halos (and subhalos) above two different mass thresholds.
The top panel of Figure~\ref{fig:Halo-Corr} includes all the halos (and subhalos) with $M_{\mathrm h} > 2.5 \times 10^{11} \msun$ which is our applied halo mass threshold for well resolved objects, and the lower panels apply a higher $M_{\mathrm h}$ threshold of $10^{12} \msun$.

All the lines in Figure~\ref{fig:Halo-Corr} show the averaged value of $\xi(r)$ which we measure from the 10 test sets, while the shaded area corresponds to $1\sigma$ standard deviation.
Comparison between the HR and SR results show that the 2-halo term $\xi^{2h}$ matches well at both redshifts. 
The SR field has a slightly suppressed $\xi^{1h}$ term compared with the HR result.
For the overall halo population (upper panels), the $\xi^{1h}$ term for SR deviates from HR by about 30\% for $z=2$ and 50\% for $z=0$ at separation distance $r \lsim 0.5 \hmpc$, which is a direct consequence of the lack of subhalo abundance in SR field.

In the lower panel of Figure~\ref{fig:Halo-Corr}, we increase the halo mass threshold and calculate $\xi (r)$ for halo (and subhalo) population above $M_{\mathrm h} > 10^{12} \msun$, for which the subhalo abundance of HR and SR in that mass range agrees better (left panel of Figure~\ref{fig:sub-halo-mf-hod}).
The resultant $\xi^{1h} (r)$ term of SR and HR differences are only at the $1\sigma$ level, in better agreement compared with $\xi^{1h} (r)$ term in the upper panel. 
Moreover, for this halo mass threshold, $\xi (r)$ has a higher normalization than before. Higher clustering is indeed expected for more massive halos.


\subsubsection{Redshift space correlations}
\label{subsubsection:halo-rsd}

\begin{figure}
\centering
  \includegraphics[width=1.0\columnwidth]{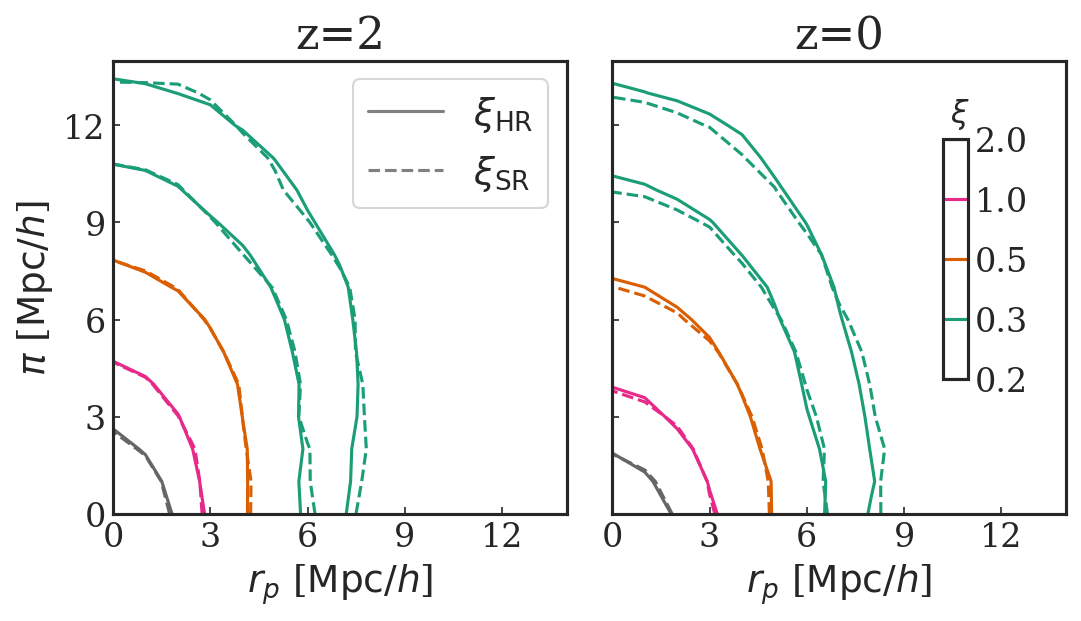}
  \caption{The contour of two-dimensional redshift space correlation function $\xi(r_p,\pi)$ for redshift-space halos in HR (solid lines) and SR (dashed lines) field at $z=2$ and $z=0$. $r_p$ is separation across the line of sight, and $\pi$ is the separation along the line of sight. 
  Here we include all the halos with $M_{\mathrm h} > 2.5 \times 10^{11} \msun$.
  Contours show lines of constant $\xi$ at $\xi=$ 2, 1, 0.5, 0.3, 0.2 respectively. The contours are calculated based on the averaged result of $\xi(r_p,\pi)$ field on the 10 test sets.}
  \label{fig:Halo-Corr-2D}
\end{figure}

\begin{figure}
\centering
  \includegraphics[width=1.0\columnwidth]{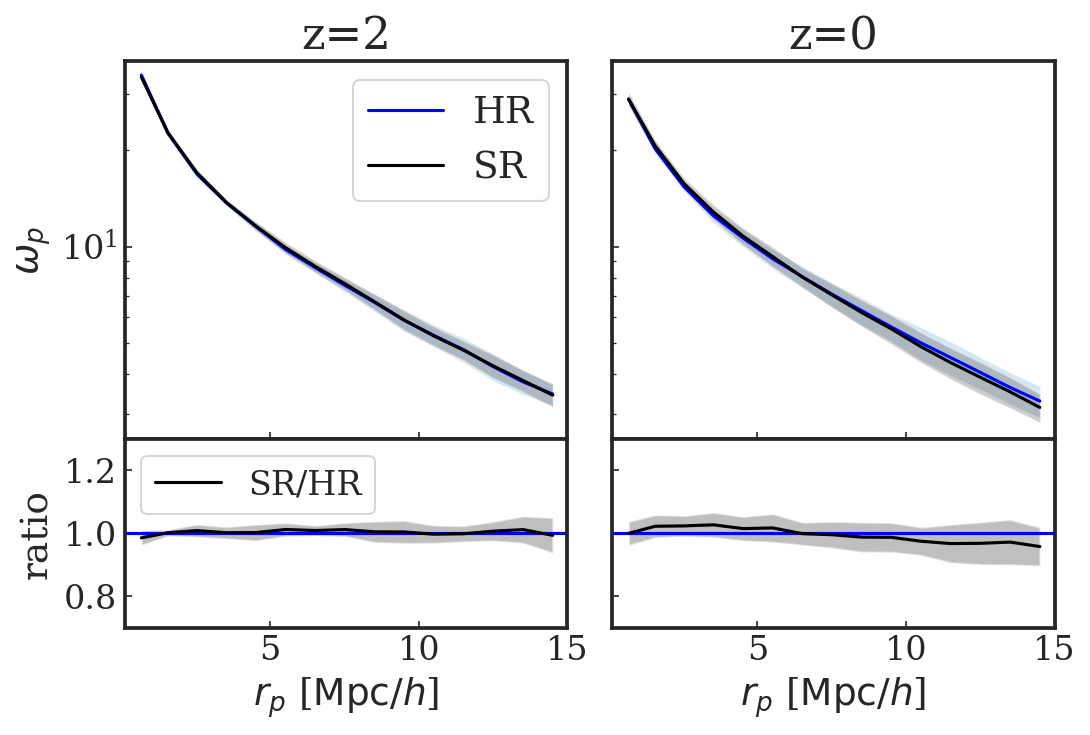}
  \caption{The projected correlation function $\omega_p(r_p)$ for redshift-space halos in HR (solid lines) and SR (dashed lines) field at $z=2$ and $z=0$. $r_p$ is separation across the line of sight. The shaded areas gives $1\sigma$ deviation measured from the 10 test sets. The lower panel gives the ratio between the SR and HR result for the test sets.}
  \label{fig:Halo-wp}
\end{figure}

Redshift space statistics encode information from the peculiar velocities that arise due to
gravitational evolution, and so are an important complementary probe of the structure formation process.
Peculiar velocities of galaxies distort large scale structures in redshift space, rendering the real space isotropic distribution of galaxies instead statistically anisotropic, and with a unique pattern which can be quantified.
In particular, redshift space distortions due to peculiar velocities enhance power along the line of sight on large scales due to the Kaiser effect \citep{Kaiser1987} and reduce power on small scales due to virialized motions
causing so-called Fingers-of-God (FOG) effect.

Figure~\ref{fig:Halo-Corr-2D} shows measurements of the 2D correlation function $\xi(r_p,\pi)$ based on the HR (solid lines) and SR (dashed lines) catalogs in redshift space at $z=2$ and $z=0$.
We account for the redshift space distortion by moving the halos along a specified line of sight according to their peculiar velocity (c.f. Equation~\ref{equation:rsd}) and measuring the resultant spatial correlation $\xi$ as a function of separation $r_p$ across and $\pi$ along the line of sight. 
The lines in Figure~\ref{fig:Halo-Corr-2D} show the contours of constant $\xi$ at $\xi=$ 2, 1, 0.5, 0.3, 0.2 respectively, calculated based on the $\xi(r_p,\pi)$ averaged over the 10 test sets.

The limited volume ($100 \hmpc$) of our training and test sets affects our ability to simulate the large-scale coherent peculiar motions which lead to Kaiser enhancement of power. 
These scales are well into the regime resolved by the LR simulation however, and so we are free to concentrate on measurements of $\xi(r_p,\pi)$ on scales $r \lsim 10 \hmpc$, where the FOG effect dominates.
The contours of $\xi(r_p,\pi)$ shown in Figure~\ref{fig:Halo-Corr-2D} are therefore elongated in the $\pi$ direction and compressed in the $r_p$ direction.
Overall, the contours of $\xi_{\mathrm{HR}}$ and $\xi_{\mathrm{SR}}$ show good agreement at $z=2$ and $z=0$, indicating that the SR field generates reasonable peculiar velocity distribution for virialized objects.
However, we do note that $\xi_{\mathrm{SR}}$ at $z=0$ is slightly less elongated in $\pi$ direction compared to $\xi_{\mathrm{HR}}$ at separation $r > 6 \hmpc$, partly due to the limited statistics at large separations in our test volume.
We will further examine the peculiar velocity distribution in the next subsection.

What is often measured in the observations, however, is the real spatial clustering of galaxies that projects the 2D correlation $\xi(r_p,\pi)$ along the $r_p$ axis.
\begin{equation}
    \omega_p(r_p) = 2 \int^{\infty}_{0} d\pi \xi(r_p,\pi)
\end{equation}
As redshift space distortions affect only the line-of-sight component of $\xi(r_p,\pi)$, integrating over the $\pi$ direction leads to a statistic $\omega_p(r_p)$ which is independent of redshift space distortions.
In Figure~\ref{fig:Halo-wp}, we plot the projected correlation function $\omega_p(r_p)$ for halos in redshift space for the HR and SR fields and give their ratio in the lower panel.
The HR and SR field measurements of $\omega_p(r_p)$ are in good agreement, with differences within the $1\sigma$ standard deviation.

\subsubsection{The pair-wise velocity distribution of halos}
\label{subsubsection: halo-velocity}

\begin{figure}
\centering
  \includegraphics[width=1.0\columnwidth]{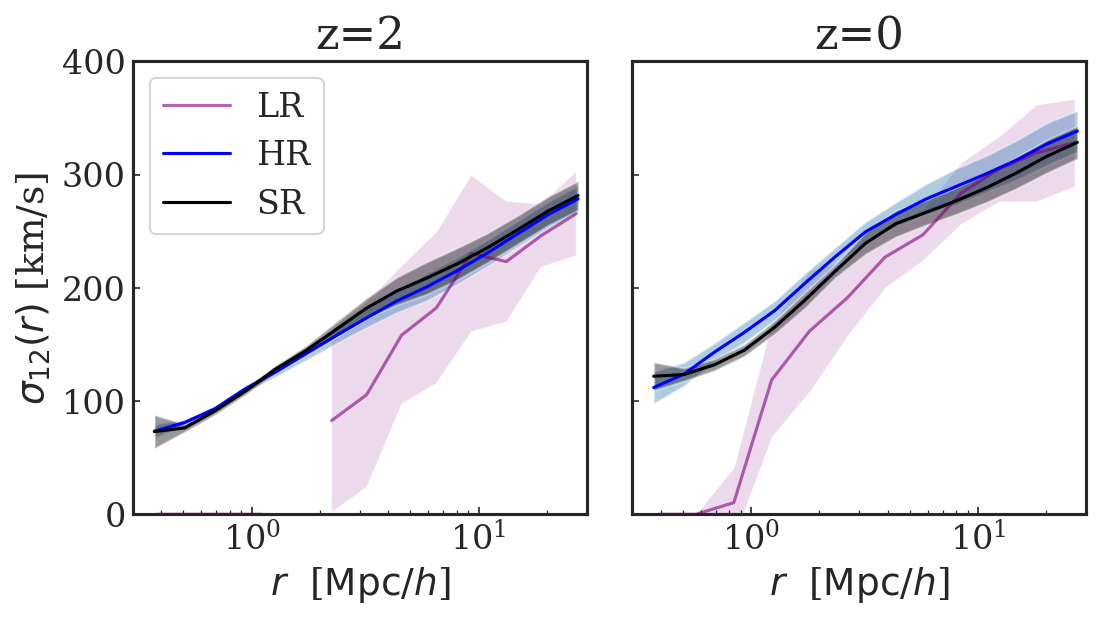}
  \caption{The pair-wise velocity distribution of the halos in the LR (purple), HR (blue) and SR (black) fields. Here $\sigma_{12}(r)$ is dispersion of the radial pairwise velocities of all halos with $M_{\mathrm{h}} > 2.5 \times 10^{11} \msun$ as a function of the distance between the halos. The shaded area shows the $1\sigma$ 
  standard deviation measured from the test sets.}
  \label{fig:pvd}
\end{figure}

The pairwise peculiar velocity dispersion of galaxies is an important measure of the structure and clustering of the universe on large scales and is used as a test of cosmological models.
As a direct validation of the peculiar velocity field of the halos, we examine the radial pair-wise velocity distribution of halos as a function of the distance between  halo pairs.

The radial pair-wise velocity $v_{12} (r)$ of halo is defined as
\begin{equation}
    v_{12}(r) = \vec{v}_{\rm 1} \cdot \vec{r}_{12}-\vec{v}_{\rm 2} \cdot \vec{r}_{12}
\end{equation}
where $\vec{v}$ is the mass center (peculiar) velocity of the halo. 
In Figure~\ref{fig:pvd}, we select all the halos with mass $M_{\mathrm{h}} > 2.5 \times 10^{11} \msun$ and give the dispersion of the radial pairwise velocity for the halo pairs separated by distance lying within the corresponding radial bin.
The lines in Figure~\ref{fig:pvd} show the averaged $\sigma_{12} (r)$ measured from the 10 test sets, and the shaded area corresponds to the $1\sigma$ standard deviation.
The velocity dispersions for the HR and SR field are good matches. 
Compared to that of HR, the averaged $\sigma_{12}$ for SR is about 3\% higher at $z=2$ and about 5\% lower at $z=0$, within $1\sigma$ standard deviation for  the different realizations.
For comparison we also show the LR result as a purple curve. The LR curve has large variations because of the limited number of halos that can be resolved. It drops off at small separations due to the missing low mass halo population.



\section{Discussion}
\label{section4:Discussion}

Applying SR techniques to cosmological simulations has also been explored by \cite{ramanah20}. In that work, the SR dark matter density fields are generated from an LR input with 2 times lower spatial resolution.
The SR model presented in our work, however, is trained to produce the full 6D phase space output of the dynamic field in the Lagrangian prescription (represented by the displacement and the velocity field of the particle distribution). 
The displacement field contains more information than the density field: different displacement fields can produce the same density field.
With the full phase space distribution, our generated SR field can be treated in the same
fashion as an output from a real $N$-body simulation with distinguishable tracer particles.
Moreover, this framework allows us to generalize the SR model to trace the time evolution of the dark matter particles, something that we plan to do in follow-up work.

In this work we have gone $8^3$ times higher in mass resolution in the SR compared to the LR simulation. While this task seems ambitious (to predict scales that are orders of magnitude below the LR input) here we have assessed quantitatively that the generated small scale structure and higher order statistics in the SR are well captured, just as they arise from complicated non-linear structure formation process in a direct $N$-body simulation. 
In terms of computing requirements, with $64^3$ particles in a $100 \, \hmpc$ volume, the LR simulations takes less than 1 core hour, easily achievable on a personal laptop. 
The corresponding $512^3$ HR simulation would take more than 2k core hours to evolve to $z=0$. 
The generation of SR field, however, takes only a few seconds ($\sim$ ten) on a single GPU (including the I/O time), a tiny fraction of the cost of the HR counterpart. 
The success with our SR models prompts us to invest further in the exploration of these techniques for cosmological hydrodynamic simulations. The high resolution required to capture baryon physics appropriately on  small (galaxy) scales and the subsequent increase in computational cost makes the application of SR techniques even more relevant for hydrodynamic simulations.
The recent, ambitious CAMEL project \citep{Villanavarro2020a}, should provide ideal training data to embark in this direction.

In this work, we focus on validating our SR model by statistically comparing the generated SR fields with the HR counterparts from real $N$-body simulations. 
Therefore, we have performed our analysis on 10 test sets with the same volume ($100 \, \hmpc$) as that of the training sets. We chose this because HR simulations of a larger volume become expensive to run.
In Paper I, however, we have shown that our SR model can be deployed to a cosmic volume of $1 \hgpc$, much larger than the $100 \, \hmpc$ training sets. 
The generated $1 \hgpc$ SR field produces good statistics for halo abundance and yields reasonable morphologies for large structures unseen in the training sets, indicating that our SR model is likely to generalize well to larger volumes.
However, we note that, in order to apply our SR model to a large cosmic volume for mock catalog generation, one may want to use larger volume simulation as training data, so that the SR model can be trained on a larger number of massive objects (e.g. $\sim 10^{15} \msun$ clusters) and produce small scale features corresponding to large modes.

In our present study, we have extended the work of Paper I into the regime of substructure of virialized objects, as well as now being able to predict the full velocity and position phase space of the SR simulation particles. There are 
many next steps, even with dark matter
only simulations, and we aim to address them in the future. These include training
a single NN using data from multiple redshifts, and training using
multiple cosmological models with different parameters. Both of these are necessary
for the SR technique to become a flexible cosmological tool. 

We have seen that the accuracy with which the SR modeling is able to
match HR results is of the order of one to ten percent, varying for different
statistics. Depending on the application (e.g., mock catalogs, rapid parameter study,
semi-analytic modeling base),  this may be sufficient, but if more accuracy is
required, we have seen through our work that changes in hyperparameters
and model architecture are able to bring improvments. Conventional N-body 
simulation methods
(e.g., \citealt{Feng2016}) also vary in accuracy, but unlike the generative methods used here, improving the accuracy while maintaining speed is more difficult.


\section{Summary and Conclusion}
\label{section5:Conclusion}

We have compared power spectra, halo mass functions, clustering, and bispectra from our recently developed super resolution models to cosmological $N$-body simulations of corresponding resolution.
Our SR task is specified from a Lagrangian prescription, and the SR model generates the full phase space displacement and velocity field represented by tracer particles which is equivalent to the output from a real $N$-body simulation.
Conditioned on the LR input, our SR model enhances the simulation resolution by generating 512 times more tracer particles, extending to the deeply non-linear regime.

In this work we perform a detailed validation of our SR model by statistically comparing the generated small scale features of the SR field with those in HR field realizations.  Specifically, we deploy the SR model on 10 test sets of $100 \, \hmpc$ volume LR-HR simulation pairs and compare the generated SR field with the HR counterparts.

We first validate our SR model by examining statistics of the entire dark matter field: we show that our generated SR fields match the power spectra of true HR realizations to within 5\%.  This level of agreement applies well below the LR resolution limit of $k \gtrsim 10 \invhmpc$.
To evaluate the non-Gaussianity of the evolved density field, a measure
of non-linear evolution, we examine the bispectrum of the generated SR field. 
 First, we measure the bispectra in the equilateral configuration to  
 examine scale dependence. Second, we measure the cross bispectra in an isosceles triangle configurations to probe the coupling of small and large scale modes in the squeezed limit.
In both cases, the SR results yield good agreement, within 10\% of the HR results at all scales down to $k \sim 10 \invhmpc$, indicating that our SR field is able
to capture higher order statistical behaviour of the evolved density field.
We also validate the SR generated velocity field by measuring the redshift space 2D powerspectra. We find that the resultant SR field performs well on scales
 beyond those possible with the LR simulation, matching the HR field with deviations which reach about 10\% on small scales.

We expect that one important application of our SR model will be generating mock catalogs from low resolution cosmological simulations.
To this end, we identify and inspect dark matter halos and their substructures from the generated SR fields.
We show that our SR model is able to generate visually authentic small-scale structures that are well beyond the resolution of the LR input, with statistically good agreement compared to that from the direct $N$-body simulations. 

We quantitatively examine the abundance of the halo and subhalo population in the test sets. The SR field yields good agreement (within 10\%) for the overall halo population down to $M_\mathrm{h} = 2.5 \times 10^{11} \msun$, while missing some subhalos in a small mass range, [$2.5 \times 10^{11} \msun$, $10^{12} \msun$]. 
This deficit of subhalo abundance is about 12\% at $z=2$ and about 40\% at $z=0$.
Nevertheless, the SR field still gives reasonable results in terms of the halo occupation distribution, halo correlations in both real and redshift space. The pairwise velocity distribution of halos matches that of the HR output with comparable scatter. The results demonstrate the potential of AI-assisted SR modeling as a powerful tool for studying small-scale galaxy formation physics in large cosmological volumes.


\section*{Acknowledgements}
This research is part of the Frontera computing project at the Texas Advanced Computing Center. Frontera is made possible by NSF award OAC-1818253.
TDM acknowledges funding from NSF ACI-1614853, NSF AST-1616168, NASA ATP 19-ATP19-0084 and 80NSSC20K0519.
TDM and RACC also acknowledge funding from NASA ATP 80NSSC18K101, and NASA ATP NNX17AK56G, and RACC, NSF AST-1909193.
SB is supported by NSF grant AST-1817256.
This work was also supported by the NSF AI Institute: Physics of the Future, NSF PHY-2020295.
The Flatiron Institute is supported by the Simons Foundation.
We also acknowledge the code packages used in this work:
The simulations for training and testing is run with \texttt{MP-Gadget} (\url{https://github.com/MP-Gadget/MP-Gadget}).
Visualization in this work is performed with open source code \texttt{gaepsi2} (\url{https://github.com/rainwoodman/gaepsi2}).
Data and catalog analysis in this work is performed with open-source
software \texttt{PyTorch}\citep{pytorch}, \texttt{nbodykit}\citep{Hand2018}, \texttt{bskit}\citep{Foreman2020bskit}, \texttt{Amiga} halo finder \citep{Knollman2009}, and \texttt{halotool}\citep{Hearin2017}.

\section*{Data Availability}

We have released our framework \texttt{map2map} to train the SR model (\url{https://github.com/eelregit/map2map}). 
It is a \texttt{PyTorch}-based general neural network framework to transform field data.
The trained SR models and the pipeline to generate the SR fields would be available on (\url{https://github.com/yueyingn/SRS-map2map}).
Training and test sets data generated in this work will be shared on reasonable request to the corresponding author.

\bibliographystyle{mnras}
\bibliography{bib.bib}

\begin{thebibliography}{}
\makeatletter
\relax
\def\mn@urlcharsother{\let\do\@makeother \do\$\do\&\do\#\do\^\do\_\do\%\do\~}
\def\mn@doi{\begingroup\mn@urlcharsother \@ifnextchar [ {\mn@doi@}
  {\mn@doi@[]}}
\def\mn@doi@[#1]#2{\def\@tempa{#1}\ifx\@tempa\@empty \href
  {http://dx.doi.org/#2} {doi:#2}\else \href {http://dx.doi.org/#2} {#1}\fi
  \endgroup}
\def\mn@eprint#1#2{\mn@eprint@#1:#2::\@nil}
\def\mn@eprint@arXiv#1{\href {http://arxiv.org/abs/#1} {{\tt arXiv:#1}}}
\def\mn@eprint@dblp#1{\href {http://dblp.uni-trier.de/rec/bibtex/#1.xml}
  {dblp:#1}}
\def\mn@eprint@#1:#2:#3:#4\@nil{\def\@tempa {#1}\def\@tempb {#2}\def\@tempc
  {#3}\ifx \@tempc \@empty \let \@tempc \@tempb \let \@tempb \@tempa \fi \ifx
  \@tempb \@empty \def\@tempb {arXiv}\fi \@ifundefined
  {mn@eprint@\@tempb}{\@tempb:\@tempc}{\expandafter \expandafter \csname
  mn@eprint@\@tempb\endcsname \expandafter{\@tempc}}}

\bibitem[\protect\citeauthoryear{Alves~de Oliveira, Li, Villaescusa-Navarro, Ho
   \& Spergel}{Alves~de Oliveira et~al.}{2020}]{Renan20}
Alves~de Oliveira R.,  Li Y.,  Villaescusa-Navarro F.,  Ho S.,   Spergel D.~N.,
   2020, NeurIPS 2020 Machine Learning and the Physical Sciences Workshop

\bibitem[\protect\citeauthoryear{{Bagla}}{{Bagla}}{2002}]{bagla02}
{Bagla} J.~S.,  2002, \mn@doi [Journal of Astrophysics and Astronomy]
  {10.1007/BF02702282}, \href
  {https://ui.adsabs.harvard.edu/abs/2002JApA...23..185B} {23, 185}

\bibitem[\protect\citeauthoryear{{Barnes} \& {Hut}}{{Barnes} \&
  {Hut}}{1986}]{barnes86}
{Barnes} J.,  {Hut} P.,  1986, \mn@doi [\nat] {10.1038/324446a0}, \href
  {https://ui.adsabs.harvard.edu/abs/1986Natur.324..446B} {324, 446}

\bibitem[\protect\citeauthoryear{{Berger} \& {Stein}}{{Berger} \&
  {Stein}}{2019}]{berger2019}
{Berger} P.,  {Stein} G.,  2019, \mn@doi [\mnras] {10.1093/mnras/sty2949},
  \href {https://ui.adsabs.harvard.edu/abs/2019MNRAS.482.2861B} {482, 2861}

\bibitem[\protect\citeauthoryear{{Bernardeau}, {Colombi}, {Gazta{\~n}aga}  \&
  {Scoccimarro}}{{Bernardeau} et~al.}{2002}]{Bernardeau2002}
{Bernardeau} F.,  {Colombi} S.,  {Gazta{\~n}aga} E.,   {Scoccimarro} R.,  2002,
  \mn@doi [\physrep] {10.1016/S0370-1573(02)00135-7}, \href
  {https://ui.adsabs.harvard.edu/abs/2002PhR...367....1B} {367, 1}

\bibitem[\protect\citeauthoryear{{Bernardini}, {Mayer}, {Reed}  \&
  {Feldmann}}{{Bernardini} et~al.}{2020}]{Bernardini2020}
{Bernardini} M.,  {Mayer} L.,  {Reed} D.,   {Feldmann} R.,  2020, \mn@doi
  [\mnras] {10.1093/mnras/staa1911}, \href
  {https://ui.adsabs.harvard.edu/abs/2020MNRAS.496.5116B} {496, 5116}

\bibitem[\protect\citeauthoryear{{Dai} \& {Seljak}}{{Dai} \&
  {Seljak}}{2021}]{Dai2021}
{Dai} B.,  {Seljak} U.,  2021, \mn@doi [Proceedings of the National Academy of
  Science] {10.1073/pnas.2020324118}, \href
  {https://ui.adsabs.harvard.edu/abs/2021PNAS..11820324D} {118, 2020324118}

\bibitem[\protect\citeauthoryear{{Feng}, {Di-Matteo}, {Croft}, {Bird},
  {Battaglia}  \& {Wilkins}}{{Feng} et~al.}{2016}]{Feng2016}
{Feng} Y.,  {Di-Matteo} T.,  {Croft} R.~A.,  {Bird} S.,  {Battaglia} N.,
  {Wilkins} S.,  2016, \mn@doi [\mnras] {10.1093/mnras/stv2484}, \href
  {http://adsabs.harvard.edu/abs/2016MNRAS.455.2778F} {455, 2778}

\bibitem[\protect\citeauthoryear{{Foreman}, {Coulton}, {Villaescusa-Navarro}
  \& {Barreira}}{{Foreman} et~al.}{2020}]{Foreman2020bskit}
{Foreman} S.,  {Coulton} W.,  {Villaescusa-Navarro} F.,   {Barreira} A.,  2020,
  \mn@doi [\mnras] {10.1093/mnras/staa2523}, \href
  {https://ui.adsabs.harvard.edu/abs/2020MNRAS.498.2887F} {498, 2887}

\bibitem[\protect\citeauthoryear{Goodfellow, Pouget-Abadie, Mirza, Xu,
  Warde-Farley, Ozair, Courville  \& Bengio}{Goodfellow
  et~al.}{2014}]{goodfellow2014generative}
Goodfellow I.,  Pouget-Abadie J.,  Mirza M.,  Xu B.,  Warde-Farley D.,  Ozair
  S.,  Courville A.,   Bengio Y.,  2014, in Advances in neural information
  processing systems. pp 2672--2680

\bibitem[\protect\citeauthoryear{{Grudi{\'c}}, {Guszejnov}, {Hopkins}, {Offner}
   \& {Faucher-Gigu{\`e}re}}{{Grudi{\'c}} et~al.}{2020}]{grudi20}
{Grudi{\'c}} M.~Y.,  {Guszejnov} D.,  {Hopkins} P.~F.,  {Offner} S. S.~R.,
  {Faucher-Gigu{\`e}re} C.-A.,  2020, arXiv e-prints, \href
  {https://ui.adsabs.harvard.edu/abs/2020arXiv201011254G} {p. arXiv:2010.11254}

\bibitem[\protect\citeauthoryear{Gulrajani, Ahmed, Arjovsky, Dumoulin  \&
  Courville}{Gulrajani et~al.}{2017}]{wgan_gp}
Gulrajani I.,  Ahmed F.,  Arjovsky M.,  Dumoulin V.,   Courville A.~C.,  2017,
  in Guyon I.,  Luxburg U.~V.,  Bengio S.,  Wallach H.,  Fergus R.,
  Vishwanathan S.,   Garnett R.,  eds, Advances in {Neural} {Information}
  {Processing} {Systems} 30. Curran Associates, Inc., pp 5767--5777, \url
  {http://papers.nips.cc/paper/7159-improved-training-of-wasserstein-gans.pdf}

\bibitem[\protect\citeauthoryear{{Hand}, {Feng}, {Beutler}, {Li}, {Modi},
  {Seljak}  \& {Slepian}}{{Hand} et~al.}{2018}]{Hand2018}
{Hand} N.,  {Feng} Y.,  {Beutler} F.,  {Li} Y.,  {Modi} C.,  {Seljak} U.,
  {Slepian} Z.,  2018, \mn@doi [\aj] {10.3847/1538-3881/aadae0}, \href
  {https://ui.adsabs.harvard.edu/abs/2018AJ....156..160H} {156, 160}

\bibitem[\protect\citeauthoryear{He, Zhang, Ren  \& Sun}{He
  et~al.}{2016}]{he2016deep}
He K.,  Zhang X.,  Ren S.,   Sun J.,  2016, in Proceedings of the IEEE
  conference on computer vision and pattern recognition. pp 770--778

\bibitem[\protect\citeauthoryear{{He}, {Li}, {Feng}, {Ho}, {Ravanbakhsh},
  {Chen}  \& {P{\'o}czos}}{{He} et~al.}{2019}]{he2019learning}
{He} S.,  {Li} Y.,  {Feng} Y.,  {Ho} S.,  {Ravanbakhsh} S.,  {Chen} W.,
  {P{\'o}czos} B.,  2019, \mn@doi [Proceedings of the National Academy of
  Science] {10.1073/pnas.1821458116}, \href
  {https://ui.adsabs.harvard.edu/abs/2019PNAS..11613825H} {116, 13825}

\bibitem[\protect\citeauthoryear{{Hearin} et~al.,}{{Hearin}
  et~al.}{2017}]{Hearin2017}
{Hearin} A.~P.,  et~al., 2017, \mn@doi [\aj] {10.3847/1538-3881/aa859f}, \href
  {https://ui.adsabs.harvard.edu/abs/2017AJ....154..190H} {154, 190}

\bibitem[\protect\citeauthoryear{{Hinshaw} et~al.,}{{Hinshaw}
  et~al.}{2013}]{hinshaw13}
{Hinshaw} G.,  et~al., 2013, \mn@doi [\apjs] {10.1088/0067-0049/208/2/19},
  \href {https://ui.adsabs.harvard.edu/abs/2013ApJS..208...19H} {208, 19}

\bibitem[\protect\citeauthoryear{Isola, Zhu, Zhou  \& Efros}{Isola
  et~al.}{2017}]{isola2017image}
Isola P.,  Zhu J.-Y.,  Zhou T.,   Efros A.~A.,  2017, in Proceedings of the
  IEEE conference on computer vision and pattern recognition. pp 1125--1134

\bibitem[\protect\citeauthoryear{{Jing}}{{Jing}}{2005}]{Jing05}
{Jing} Y.~P.,  2005, \mn@doi [\apj] {10.1086/427087}, \href
  {https://ui.adsabs.harvard.edu/abs/2005ApJ...620..559J} {620, 559}

\bibitem[\protect\citeauthoryear{{Kaiser}}{{Kaiser}}{1987}]{Kaiser1987}
{Kaiser} N.,  1987, \mn@doi [\mnras] {10.1093/mnras/227.1.1}, \href
  {https://ui.adsabs.harvard.edu/abs/1987MNRAS.227....1K} {227, 1}

\bibitem[\protect\citeauthoryear{{Khim} et~al.,}{{Khim} et~al.}{2020}]{khim20}
{Khim} D.~J.,  et~al., 2020, \mn@doi [\apj] {10.3847/1538-4357/ab88a9}, \href
  {https://ui.adsabs.harvard.edu/abs/2020ApJ...894..106K} {894, 106}

\bibitem[\protect\citeauthoryear{{Knollmann} \& {Knebe}}{{Knollmann} \&
  {Knebe}}{2009}]{Knollman2009}
{Knollmann} S.~R.,  {Knebe} A.,  2009, \mn@doi [\apjs]
  {10.1088/0067-0049/182/2/608}, \href
  {https://ui.adsabs.harvard.edu/abs/2009ApJS..182..608K} {182, 608}

\bibitem[\protect\citeauthoryear{{Kodi Ramanah}, {Charnock}  \& {Lavaux}}{{Kodi
  Ramanah} et~al.}{2019}]{ramanah2019painting}
{Kodi Ramanah} D.,  {Charnock} T.,   {Lavaux} G.,  2019, \mn@doi [\prd]
  {10.1103/PhysRevD.100.043515}, \href
  {https://ui.adsabs.harvard.edu/abs/2019PhRvD.100d3515K} {100, 043515}

\bibitem[\protect\citeauthoryear{{Kodi Ramanah}, {Charnock},
  {Villaescusa-Navarro}  \& {Wandelt}}{{Kodi Ramanah}
  et~al.}{2020}]{KodiRamanah2020}
{Kodi Ramanah} D.,  {Charnock} T.,  {Villaescusa-Navarro} F.,   {Wandelt}
  B.~D.,  2020, \mn@doi [\mnras] {10.1093/mnras/staa1428}, \href
  {https://ui.adsabs.harvard.edu/abs/2020MNRAS.495.4227K} {495, 4227}

\bibitem[\protect\citeauthoryear{{Landy} \& {Szalay}}{{Landy} \&
  {Szalay}}{1993}]{Landy1993}
{Landy} S.~D.,  {Szalay} A.~S.,  1993, \mn@doi [\apj] {10.1086/172900}, \href
  {https://ui.adsabs.harvard.edu/abs/1993ApJ...412...64L} {412, 64}

\bibitem[\protect\citeauthoryear{{Ledig} et~al.,}{{Ledig}
  et~al.}{2016}]{Ledig2016}
{Ledig} C.,  et~al., 2016, arXiv e-prints, \href
  {https://ui.adsabs.harvard.edu/abs/2016arXiv160904802L} {p. arXiv:1609.04802}

\bibitem[\protect\citeauthoryear{{Li}, {Ni}, {Croft}, {Di Matteo}, {Bird}  \&
  {Feng}}{{Li} et~al.}{2020}]{Li2021}
{Li} Y.,  {Ni} Y.,  {Croft} R. A.~C.,  {Di Matteo} T.,  {Bird} S.,   {Feng} Y.,
   2020, arXiv e-prints, \href
  {https://ui.adsabs.harvard.edu/abs/2020arXiv201006608L} {p. arXiv:2010.06608}

\bibitem[\protect\citeauthoryear{{Modi}, {Feng}  \& {Seljak}}{{Modi}
  et~al.}{2018}]{Modi2018}
{Modi} C.,  {Feng} Y.,   {Seljak} U.,  2018, \mn@doi [\jcap]
  {10.1088/1475-7516/2018/10/028}, \href
  {https://ui.adsabs.harvard.edu/abs/2018JCAP...10..028M} {2018, 028}

\bibitem[\protect\citeauthoryear{{Nagamine}}{{Nagamine}}{2018}]{nagamine18}
{Nagamine} K.,  2018, {The Encyclopedia of Cosmology. Volume 2: Numerical
  Simulations in Cosmology}, \mn@doi{10.1142/9496-vol2.
}

\bibitem[\protect\citeauthoryear{{Nelson} et~al.,}{{Nelson}
  et~al.}{2020}]{nelson20}
{Nelson} D.,  et~al., 2020, \mn@doi [\mnras] {10.1093/mnras/staa2419}, \href
  {https://ui.adsabs.harvard.edu/abs/2020MNRAS.498.2391N} {498, 2391}

\bibitem[\protect\citeauthoryear{Paszke et~al.,}{Paszke et~al.}{2019}]{pytorch}
Paszke A.,  et~al., 2019, in Wallach H.,  Larochelle H.,  Beygelzimer A.,
  d\textquotesingle Alch\'{e}-Buc F.,  Fox E.,   Garnett R.,  eds, , Advances
  in Neural Information Processing Systems 32.
Curran Associates, Inc., pp 8024--8035

\bibitem[\protect\citeauthoryear{{Percival}, {Samushia}, {Ross}, {Shapiro}  \&
  {Raccanelli}}{{Percival} et~al.}{2011}]{percival11}
{Percival} W.~J.,  {Samushia} L.,  {Ross} A.~J.,  {Shapiro} C.,   {Raccanelli}
  A.,  2011, \mn@doi [Philosophical Transactions of the Royal Society of London
  Series A] {10.1098/rsta.2011.0370}, \href
  {https://ui.adsabs.harvard.edu/abs/2011RSPTA.369.5058P} {369, 5058}

\bibitem[\protect\citeauthoryear{{Perraudin}, {Srivastava}, {Lucchi},
  {Kacprzak}, {Hofmann}  \& {R{\'e}fr{\'e}gier}}{{Perraudin}
  et~al.}{2019}]{Perraudin2019}
{Perraudin} N.,  {Srivastava} A.,  {Lucchi} A.,  {Kacprzak} T.,  {Hofmann} T.,
   {R{\'e}fr{\'e}gier} A.,  2019, \mn@doi [Computational Astrophysics and
  Cosmology] {10.1186/s40668-019-0032-1}, \href
  {https://ui.adsabs.harvard.edu/abs/2019ComAC...6....5P} {6, 5}

\bibitem[\protect\citeauthoryear{{Perraudin}, {Marcon}, {Lucchi}  \&
  {Kacprzak}}{{Perraudin} et~al.}{2020}]{Perraudin2020}
{Perraudin} N.,  {Marcon} S.,  {Lucchi} A.,   {Kacprzak} T.,  2020, arXiv
  e-prints, \href {https://ui.adsabs.harvard.edu/abs/2020arXiv200408139P} {p.
  arXiv:2004.08139}

\bibitem[\protect\citeauthoryear{{Porter}}{{Porter}}{1985}]{porter85}
{Porter} D.~H.,  1985, PhD thesis, California Univ., Berkeley.

\bibitem[\protect\citeauthoryear{Ramanah, {Charnock}, {Villaescusa-Navarro}  \&
  {Wandelt}}{Ramanah et~al.}{2020}]{ramanah20}
Ramanah D.~K.,  {Charnock} T.,  {Villaescusa-Navarro} F.,   {Wandelt} B.~D.,
  2020, \mn@doi [\mnras] {10.1093/mnras/staa1428}, \href
  {https://ui.adsabs.harvard.edu/abs/2020MNRAS.495.4227K} {495, 4227}

\bibitem[\protect\citeauthoryear{{Rodrigues} et~al.,}{{Rodrigues}
  et~al.}{2016}]{rodrigues16}
{Rodrigues} M.,  et~al., 2016, \mn@doi [\aap] {10.1051/0004-6361/201527836},
  \href {https://ui.adsabs.harvard.edu/abs/2016A&A...590A..18R} {590, A18}

\bibitem[\protect\citeauthoryear{{Rodr{\'\i}guez}, {Kacprzak}, {Lucchi},
  {Amara}, {Sgier}, {Fluri}, {Hofmann}  \&
  {R{\'e}fr{\'e}gier}}{{Rodr{\'\i}guez} et~al.}{2018}]{Rodrguez2018}
{Rodr{\'\i}guez} A.~C.,  {Kacprzak} T.,  {Lucchi} A.,  {Amara} A.,  {Sgier} R.,
   {Fluri} J.,  {Hofmann} T.,   {R{\'e}fr{\'e}gier} A.,  2018, \mn@doi
  [Computational Astrophysics and Cosmology] {10.1186/s40668-018-0026-4}, \href
  {https://ui.adsabs.harvard.edu/abs/2018ComAC...5....4R} {5, 4}

\bibitem[\protect\citeauthoryear{{Salmon}}{{Salmon}}{1991}]{salmon91}
{Salmon} J.~K.,  1991, PhD thesis, California Institute of Technology,
  Pasadena.

\bibitem[\protect\citeauthoryear{{Scoccimarro}}{{Scoccimarro}}{2000}]{Scoccimarro2000}
{Scoccimarro} R.,  2000, \mn@doi [\apj] {10.1086/317248}, \href
  {https://ui.adsabs.harvard.edu/abs/2000ApJ...544..597S} {544, 597}

\bibitem[\protect\citeauthoryear{Shi, Caballero, Husz{\'a}r, Totz, Aitken,
  Bishop, Rueckert  \& Wang}{Shi et~al.}{2016}]{shi2016real}
Shi W.,  Caballero J.,  Husz{\'a}r F.,  Totz J.,  Aitken A.~P.,  Bishop R.,
  Rueckert D.,   Wang Z.,  2016, in Proceedings of the IEEE conference on
  computer vision and pattern recognition. pp 1874--1883

\bibitem[\protect\citeauthoryear{{Tr{\"o}ster}, {Ferguson},
  {Harnois-D{\'e}raps}  \& {McCarthy}}{{Tr{\"o}ster}
  et~al.}{2019}]{Troster2019}
{Tr{\"o}ster} T.,  {Ferguson} C.,  {Harnois-D{\'e}raps} J.,   {McCarthy} I.~G.,
   2019, \mn@doi [\mnras] {10.1093/mnrasl/slz075}, \href
  {https://ui.adsabs.harvard.edu/abs/2019MNRAS.487L..24T} {487, L24}

\bibitem[\protect\citeauthoryear{{Villaescusa-Navarro}
  et~al.,}{{Villaescusa-Navarro} et~al.}{2020a}]{Villanavarro2020b}
{Villaescusa-Navarro} F.,  et~al., 2020a, arXiv e-prints, \href
  {https://ui.adsabs.harvard.edu/abs/2020arXiv201000619V} {p. arXiv:2010.00619}

\bibitem[\protect\citeauthoryear{{Villaescusa-Navarro}, {Wandelt},
  {Angl{\'e}s-Alc{\'a}zar}, {Genel}, {Zorrilla Mantilla}, {Ho}  \&
  {Spergel}}{{Villaescusa-Navarro} et~al.}{2020b}]{Villanavarro2020a}
{Villaescusa-Navarro} F.,  {Wandelt} B.~D.,  {Angl{\'e}s-Alc{\'a}zar} D.,
  {Genel} S.,  {Zorrilla Mantilla} J.~M.,  {Ho} S.,   {Spergel} D.~N.,  2020b,
  arXiv e-prints, \href {https://ui.adsabs.harvard.edu/abs/2020arXiv201105992V}
  {p. arXiv:2011.05992}

\bibitem[\protect\citeauthoryear{{Vogelsberger}, {Marinacci}, {Torrey}  \&
  {Puchwein}}{{Vogelsberger} et~al.}{2020}]{vogelsberger20}
{Vogelsberger} M.,  {Marinacci} F.,  {Torrey} P.,   {Puchwein} E.,  2020,
  \mn@doi [Nature Reviews Physics] {10.1038/s42254-019-0127-2}, \href
  {https://ui.adsabs.harvard.edu/abs/2020NatRP...2...42V/abstract} {2, 42}

\bibitem[\protect\citeauthoryear{{Wadekar}, {Villaescusa-Navarro}, {Ho}  \&
  {Perreault-Levasseur}}{{Wadekar} et~al.}{2020}]{wadekar2020hinet}
{Wadekar} D.,  {Villaescusa-Navarro} F.,  {Ho} S.,   {Perreault-Levasseur} L.,
  2020, arXiv e-prints, \href
  {https://ui.adsabs.harvard.edu/abs/2020arXiv200710340W} {p. arXiv:2007.10340}

\bibitem[\protect\citeauthoryear{{Walsh} \& {Levison}}{{Walsh} \&
  {Levison}}{2019}]{walsh19}
{Walsh} K.~J.,  {Levison} H.~F.,  2019, \mn@doi [\icarus]
  {10.1016/j.icarus.2019.03.031}, \href
  {https://ui.adsabs.harvard.edu/abs/2019Icar..329...88W} {329, 88}

\bibitem[\protect\citeauthoryear{{Wang}, {Chen}  \& {Hoi}}{{Wang}
  et~al.}{2019}]{wang2020}
{Wang} Z.,  {Chen} J.,   {Hoi} S. C.~H.,  2019, arXiv e-prints, \href
  {https://ui.adsabs.harvard.edu/abs/2019arXiv190206068W} {p. arXiv:1902.06068}

\bibitem[\protect\citeauthoryear{Yue, Shen, Li, Yuan, Zhang  \& Zhang}{Yue
  et~al.}{2016}]{Yue2016}
Yue L.,  Shen H.,  Li J.,  Yuan Q.,  Zhang H.,   Zhang L.,  2016, \mn@doi
  [Signal Processing] {10.1016/j.sigpro.2016.05.002}, 128

\bibitem[\protect\citeauthoryear{{Zhang}, {Wang}, {Zhang}, {Sun}, {He},
  {Contardo}, {Villaescusa-Navarro}  \& {Ho}}{{Zhang} et~al.}{2019}]{zhang2019}
{Zhang} X.,  {Wang} Y.,  {Zhang} W.,  {Sun} Y.,  {He} S.,  {Contardo} G.,
  {Villaescusa-Navarro} F.,   {Ho} S.,  2019, arXiv e-prints, \href
  {https://ui.adsabs.harvard.edu/abs/2019arXiv190205965Z} {p. arXiv:1902.05965}

\makeatother
\end{thebibliography}

\end{document}